\documentclass[fleqn,usenatbib]{mnras}

\usepackage[T1]{fontenc}

\DeclareRobustCommand{\VAN}[3]{#2}
\let\VANthebibliography\thebibliography
\def\thebibliography{\DeclareRobustCommand{\VAN}[3]{##3}\VANthebibliography}

\usepackage{graphicx}	% Including figure files
\usepackage{amsmath}	% Advanced maths commands
\usepackage{amssymb}	% Extra maths symbols

\usepackage{graphicx}	% Including figure files
\usepackage{amsmath}	% Advanced maths commands
\usepackage{amssymb}	% Extra maths symbols
\usepackage[normalem]{ulem} %strike through text

\usepackage{hyperref}

\newcommand{\feiil}{[Fe\,{\sc ii}]1.2570\,$\mu$m}
\newcommand{\feii}{[Fe\,{\sc ii}]}
\newcommand{\pii}{[P\,{\sc ii}]}
\newcommand{\piil}{[P\,{\sc ii}]1.1886\,$\mu$m}
\newcommand{\hml}{H$_2$\,2.1218\,$\mu$m}
\newcommand{\hm}{H$_2$}
\newcommand{\brg}{Br$\gamma$}
\newcommand{\pab}{Pa$\beta$}
\newcommand{\sixl}{[S\,{\sc ix}]1.2523\,$\mu$m}
\newcommand{\six}{[S\,{\sc ix}]}

\title[Gemini NIFS AGN survey - IV. Excitation]{Gemini NIFS survey of feeding and feedback in nearby Active Galaxies - IV. Excitation}

\author[R. A. Riffel et al.]{Rogemar A. Riffel,$^{1}$\thanks{E-mail:
rogemar@ufsm.br}
 Marina Bianchin,$^{1}$ 
 Rog\'erio Riffel,$^{2}$
 Thaisa Storchi-Bergmann,$^{2}$
\newauthor 
 Astor J. Sch\"onell,$^{3}$
Luis Gabriel Dahmer-Hahn,$^{4}$
Natacha Z. Dametto,$^{5}$
Marlon R. Diniz$^{1}$\\
$^{1}$Universidade Federal de Santa Maria, Departamento de F\'isica, Centro de Ci\^encias Naturais e Exatas, 97105-900,
Santa Maria, RS, Brazil\\
$^{3}$Instituto de F\'isica, Universidade Federal do Rio Grande do Sul, Av. Bento Gon\c calves 9500, 91501-970, 
Porto Alegre, RS, Brazil \\
${^3}$ Instituto Federal de Educa\c c\~ao, Ci\^encia e Tecnologia Farroupilha, BR287, km 360, Estrada do Chapad\~ao, 97760-000, 
Jaguari - RS, Brazil\\
$^{4}$Laborat\'orio Nacional de Astrof\'isica. Rua dos Estados Unidos, 154, CEP 37504-364 Itajub\'a, MG, Brazil \\
$^{5}$Centro de Astronom\'ia (CITEVA), Universidad de Antofagasta, Avenida Angamos 601, Antofagasta, Chile\\
}
\date{Accepted XXX. Received YYY; in original form ZZZ}

\pubyear{2020}

\begin{document}
\label{firstpage}
\pagerange{\pageref{firstpage}--\pageref{lastpage}}
\maketitle

% Abstract of the paper
\begin{abstract}
The near-infrared spectra of Active Galactic Nuclei (AGN) present emission lines of different atomic and molecular species.
The mechanisms involved in the origin of these emission lines in AGN are still not fully understood. 
We use J and K band integral field spectra of six luminous ($43.1<\log L_{\rm bol}/({\rm erg\,s^{-1}})<44.4$) Seyfert galaxies (NGC\,788, Mrk\,607, NGC\,3227, NGC\,3516, NGC\,5506 and NGC\,5899) in the Local Universe ($0.0039<z<0.0136$) to investigate the gas excitation within the inner 100-300\,pc radius of the galaxies at spatial resolutions of a few tens of parsecs. 
%We find a good correlation between H$_2$\,2.1218\,$\mu$m/Br$\gamma$ and [Fe\,{\sc ii}]\,1.2570\,$\mu$m/Pa$\beta$, well reproduced by $\log \left({\rm \frac{[Fe\,II]}{Pa\beta}}\right) = {\rm(0.65\pm0.026)\times\log \left(\frac{H_2}{Br\gamma}\right) - (0.12\pm0.013)}$. 
In all galaxies, the H$_2$ emission originates from thermal processes with excitation temperatures in the range 2400--5200\,K.  In the high line ratio (HLR) region of the H$_2$/Br$\gamma$ vs. [Fe\,{\sc ii}]/Pa$\beta$ diagnostic diagram, which includes 29\,\% of the spaxels, shocks are the main excitation mechanism, as indicated by the correlation between the line widths and line ratios. In the AGN region of the diagram (64\,\% of the spaxels) the H$_2$ emission is due to the AGN radiation. The \feii\ emission is produced by a combination of photoionisation by the AGN radiation and shocks in five galaxies and is dominated by photoionisation in NGC\,788.  The \sixl\ coronal emission line is present in all galaxies, and its flux distributions are extended from 80 to 185\,pc from the galaxy nuclei, except for NGC\,5899, in which this line is detected only in the integrated spectrum. 
%The origin of the highly ionised gas is mainly due to photoionisation by the central AGN, but shocks cannot be ruled out for the most extended emission.   

\end{abstract}


%%%%%%%%%%%%%%%%% BODY OF PAPER %%%%%%%%%%%%%%%%%%%%%%%%%%%%%%%%%%%%%%%%%%%%%%

\section{Introduction}

The \feii\ and molecular hydrogen emission lines are among the most prominent spectral features in the near-infrared (near-IR) spectra of active  galactic nuclei (AGN), but which excitation mechanisms are responsible for their emission in AGN are still in debate \citep[e.g.][]{ardila04,ardila05,rogerio06,dors12,rogerio13,rogerio19,lamperti17}. Such lines can be produced by the AGN or young stars radiation field or by shocks, as for example from supernovae explosions or interaction of a radio jet with the ambient gas. 

The H$_2$ molecule can be excited by (i) fluorescence through absorption of soft-ultraviolet photons (912--1108 \AA) in the Lyman and Werner bands, existing both in star-forming regions and surrounding AGN \citep{black87},  (ii) shocks \citep{hollenbach89}, e.g. the interaction of a radio jet with the interstellar medium \citep[e.g.]{rogemar_eso428,rogemar_n5929} or from supernovae explosions \citep[][and references therein]{larkin98}, (iii) X-ray illumination by a central AGN \citep{draine90,maloney96} or  (iv) UV radiation in  dense clouds, with densities in the range 10$^4$ and 10$^5$ cm$^{-3}$ \citep{Sternberg89,davies03} -- see also \citet{mouri94} for a discussion about the origin of the H$_2$ emission in Seyfert and Starburst galaxies.  UV fluorescence is usually regarded as a non-thermal excitation process, while shocks and X-ray/UV heating are referred to as thermal processes. Thermal and non-thermal processes produce distinct relative intensities between emission lines of H$_2$, which can be used to discriminate the dominant excitation mechanism. Single aperture \citep[e.g.][]{reunanen02,ardila04,ardila05,rogerio13} and spatially resolved  \citep[e.g.][]{davies05,sbN4151Exc,mazzalay13,rogemar_n1068,rogemar_N1275} observations of AGN suggest that thermal processes dominate the production of the nuclear H$_2$ emission in Seyfert galaxies. Near-IR integral field spectroscopy (IFS) provides the spatial distribution and kinematics of the emission lines, which can be used to kinematically identify signatures of shocks (e.g. increased line widths associated to a radio jet or outflow), thus making it possible to discriminate between X-rays and shocks \citep[e.g.][]{rogemarMrk1066exc,rogemar_n5929}. 

The [Fe\,{\sc ii}] near-IR emission lines are produced in the partially ionised gas phase and [Fe\,{\sc ii}]1.2570$\mu$m/Pa$\beta$ ratio is controlled by the ratio between the volumes of the partially to fully ionised gas \citep{mouri90,mouri93,ardila05,rogemarN4051,sbN4151Exc}. The partially ionised zones in the central region of galaxies originate from X-ray emission \citep{simpson96} in AGN or shocks due to the interaction of radio jets and gas outflows with the ambient clouds \citep{forbes93}. However, to determine whether the [Fe\,{\sc ii}] emission in AGN is due to shocks or X-ray excitation, a detailed mapping of the emission-line distributions is required, which can be done by analysing IFS data of spatially-resolved outflow structures in nearby AGN or more powerful distant nuclei. Even with many works investigating the excitation processes of the H$_2$ and [Fe\,{\sc ii}], it is still unclear if they have a common origin or are produced by a combination of distinct mechanisms. 

Some studies have used near-IR IFS to map the [Fe\,{\sc ii}] and H$_2$ distribution and kinematics in local Seyfert galaxies at scales from few tens to few hundreds of parsecs \citep[e.g.][]{rogemar_eso428,rogemarMrk1066exc,rogemar_n1068,rogemar_n5929_let,rogemar_N1275,durre18,may17,may20}. The results reached by these studies suggest the H$_2$ and [Fe\,{\sc ii}] present distinct flux distributions and kinematics, with the former more restricted to the galactic plane and the latter extending to higher latitudes above it and often presenting signatures of outflows.  However, the studies above are based on single objects. A detailed study on the H$_2$ and [Fe\,{\sc ii}] emission is performed by \citet{colina15}, who investigated the gas excitation structure in a sample composed of  10 luminous infrared galaxies (LIRGs), 5 Seyfert galaxies and 7 Star-Forming galaxies, using near-IR IFS.  They define new areas in the [Fe\,{\sc ii}]1.6440\,$\mu$m/Br$\gamma$ -- H$_2$2.1218\,$\mu$m/Br$\gamma$ diagnostic plane -- previously based on single aperture measurements \citep[e.g.][]{larkin98,reunanen02,rogerio06,rogerio13},  for the compact, high surface brightness regions dominated by AGN, young ionising stars, and supernovae explosions. Although, \citet{colina15} define typical ranges of AGN line ratios, their work is not aimed to investigate in details the excitation mechanisms of the near-IR emission lines in AGN hosts. 
 
Besides the molecular and low ionisation gas, the near-IR spectra includes coronal emission lines (CLs) from highly ionisation  gas (with typical ionisation potentials larger than 100\,eV), such as [S\,{\sc ix}]1.2523\,$\mu$m, which  has an ionisation potential of 328.2\,eV. The origin of the CLs in AGN is still not clear. These lines are usually broader than those from low ionisation gas and blueshifted relative to the stellar rest frame, indicating they originate in the outflowing gas from the outer portions of the Broad Line Region \citep{ardila02,ms11}.  This is further supported by photoionisation models, which suggest that the nuclear coronal emission in AGN originates in the inner edge of the dusty torus \citep{glidden16} and by the fact that the luminosity of coronal lines is correlated with the X-ray luminosity \citep{ardila11}. However, photoionisation models cannot reproduce the extended coronal gas emission on scales of a few hundred of parsecs, where the coronal lines originate from shocks due to the interaction of a radio jet or outflows with the interestellar medium \citep[e.g.][]{ardila20}.

In this study, we use near-IR IFS of a sample of six luminous Seyfert galaxies (NGC\,788, Mrk\,607, NGC\,3227, NGC\,3516, NGC\,5506 and NGC\,5899) to investigate the origin of the [Fe\,{\sc ii}] and H$_2$ emission on scales of $\sim$100\,pc and map the spatial extent of the coronal emission lines. We use the same sample previously discussed in \citet{astor19}, where we present the data, emission-line flux, surface gas mass distributions and kinematic maps. There, we found that H$_2$ emission originates mostly from gas rotating in the galaxy plane, while the ionised gas shows signatures of outflows. In addition, the amount of gas available in the inner kpc of these galaxies is, at least, 100 times larger than needed to power the central AGN. This paper is organized as follows. Section \ref{sec:sample} presents information about the sample, data reduction and measurements, Section \ref{sec:results} presents the emission-line flux distributions and emission-line ratio diagnostic diagrams, while the discussion is presented in Section \ref{sec:discussion} and  our conclusions in Section \ref{sec:conclusions}. We use a $h=0.7, \Omega_m=0.3, \Omega_{\Lambda}=0.7$ cosmology throughout this paper.

\section{The sample, data and measurements}
\label{sec:sample}

Our sample is composed of 6 nearby and luminous Seyfert galaxies. They are part of a sample of 29 galaxies from the Gemini NIFS survey of feeding and feedback processes in nearby active galaxies \citep{rogemar_sample}. The objects on the sample were selected based on the following criteria: hard X-ray (14--195 keV)  luminosities $L_{\rm X} \geq 10^{41.5}$ erg\,s$^{-1}$, redshifts $z\leq0.015$ and extended [O\,{\sc iii}] emission previously reported in the literature. Nine galaxies from this sample have not yet been observed, mainly due to failures in the ALTtitude conjugate Adaptive optics for the InfraRed (ALTAIR)  system. Among the already observed galaxies in both J and K bands using adaptive optics, we select for this study the six objects with no previous detailed analysis on the gas excitation: NGC\,788, Mrk\,607, NGC\,3227, NGC\,3516, NGC\,5506 and NGC\,5899.   The six galaxies studied here are the same previously studied in \citet{astor19}, where we focused on the presentation of data, flux and gas mass distribution.  
They are  
%were selected for this study because previous works on them were not aimed at investigating the gas excitation. They are 
among the 50\% more luminous objects of the \citet{rogemar_sample} sample. 
The galaxies of our sample have  ${42.36}\leq \log L_{\rm X}/{\rm erg\, s^{-1}} \leq {43.51}$ \citep{BAT105} -- which corresponds to bolometric luminosites in the range  ${43.44}\leq \log L_{\rm bol}/{\rm erg\, s^{-1}} \leq {44.83}$, using the relation between $L_{\rm X}$ and $L_{\rm bol}$ presented in \citet{ichikawa17} -- and redshifts in the range  $0.004\leq z \leq 0.014$ (corresponding to 17--60 Mpc).

The J and K band observations were performed with the Gemini Near-infrared Integral Field Spectrograph \citep[NIFS,][]{mcgregor03}, which has a field of view of 3\,arcsec$\times$3\,arcsec and angular sampling of  0.103$\times$0.042\,arcsec$^2$. NIFS was coupled to the ALTtitude conjugate Adaptive optics for the InfraRed (ALTAIR) system and the resulting angular resolutions are in the range 0.12--0.18\,arcsec, corresponding to a few tens of parsecs.

We follow the standard procedure to reduce the raw data using the {\sc gemini iraf} package, including the trimming of the images, flat-fielding, cosmic ray rejection, sky subtraction, wavelength and s-distortion calibrations, removal of the telluric features, flux calibration and construction of the datacubes for each individual exposure on each science target at an angular sampling of 0.05\,arcsec$\times$0.05\,arcsec. The final datacubes are obtained by median combining the individual datacubes in each band using the peak of the continuum emission as reference. More details about the sample, instrument, observations and data reduction are presented in previous papers \citep{rogemar_stellar,rogemar_sample,astor19}.

\begin{figure}
    \centering
    \includegraphics[width=0.23\textwidth]{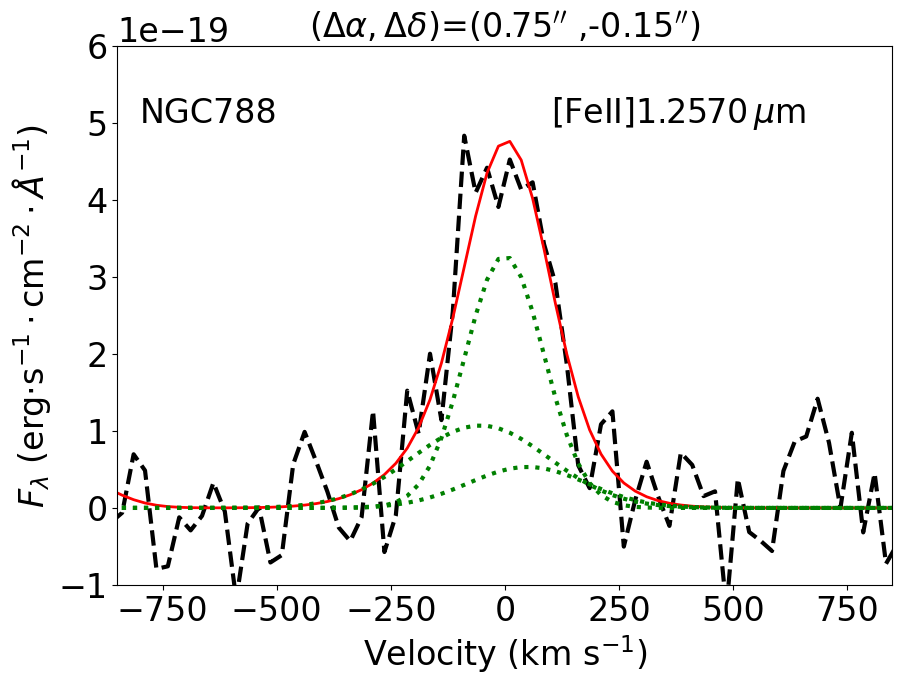} 
   \includegraphics[width=0.23\textwidth]{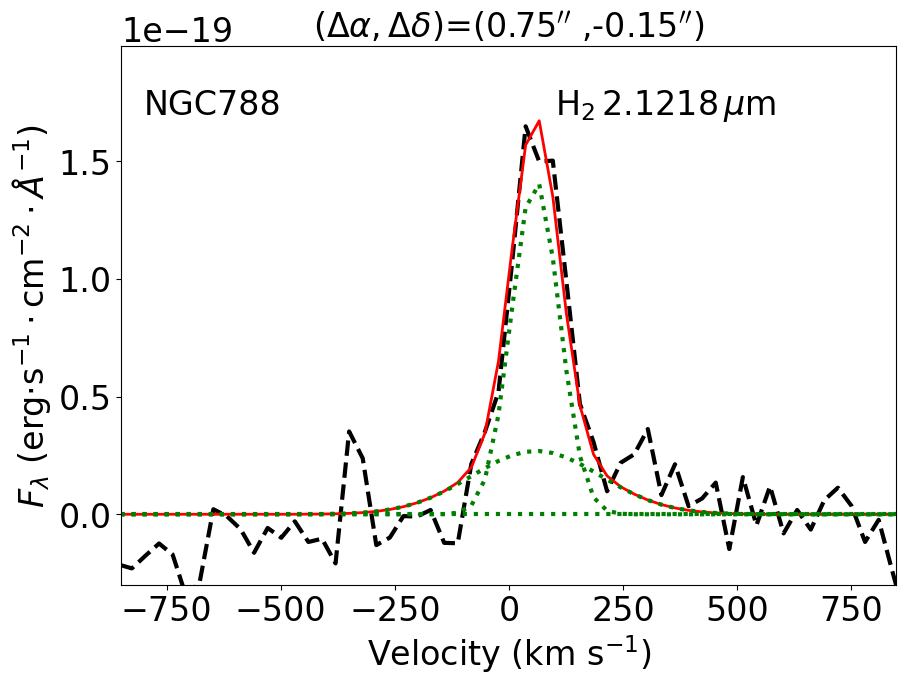}    
    \includegraphics[width=0.23\textwidth]{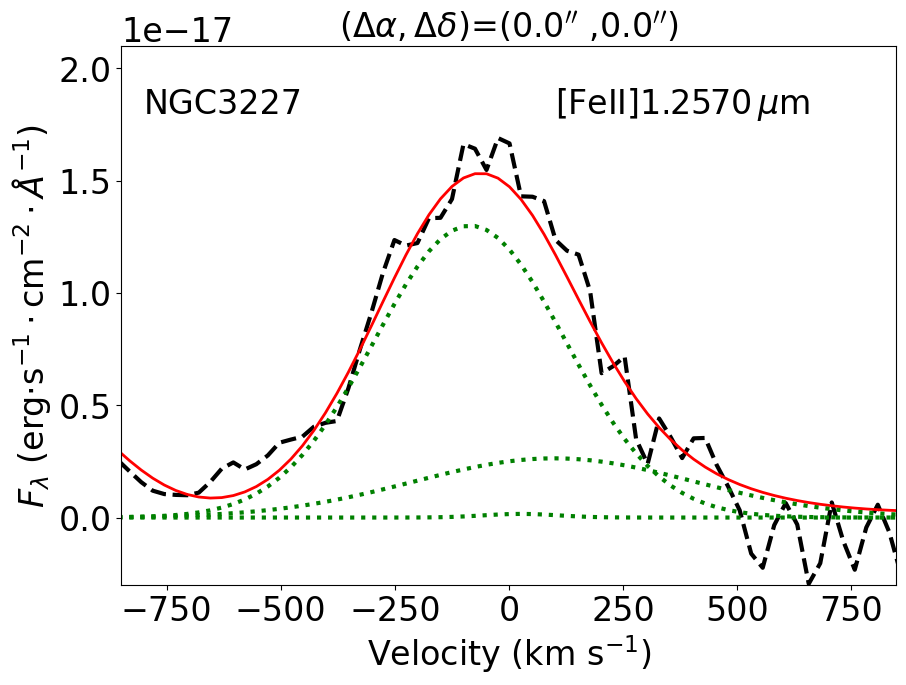} 
   \includegraphics[width=0.23\textwidth]{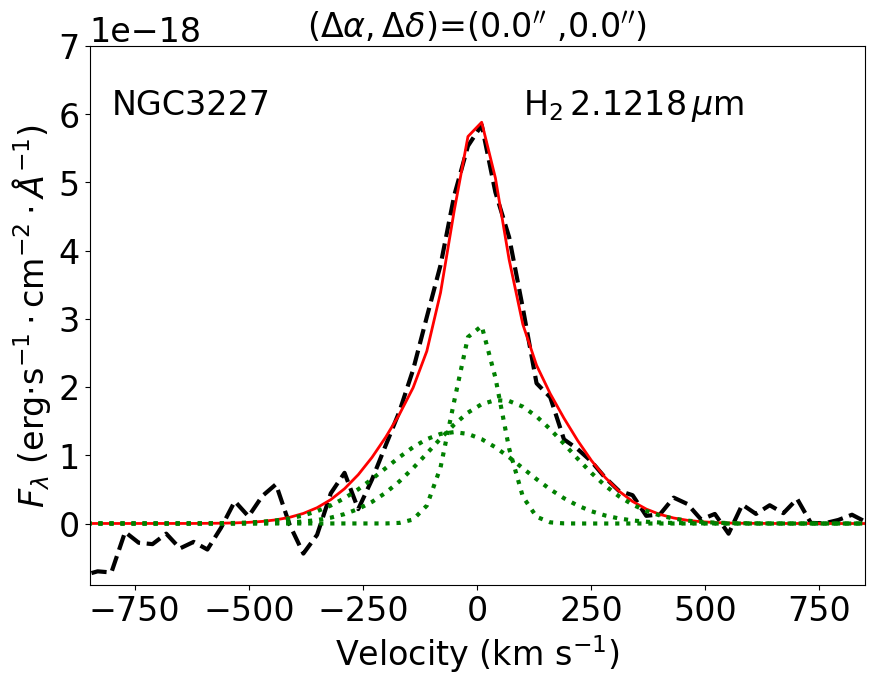}    
    \includegraphics[width=0.23\textwidth]{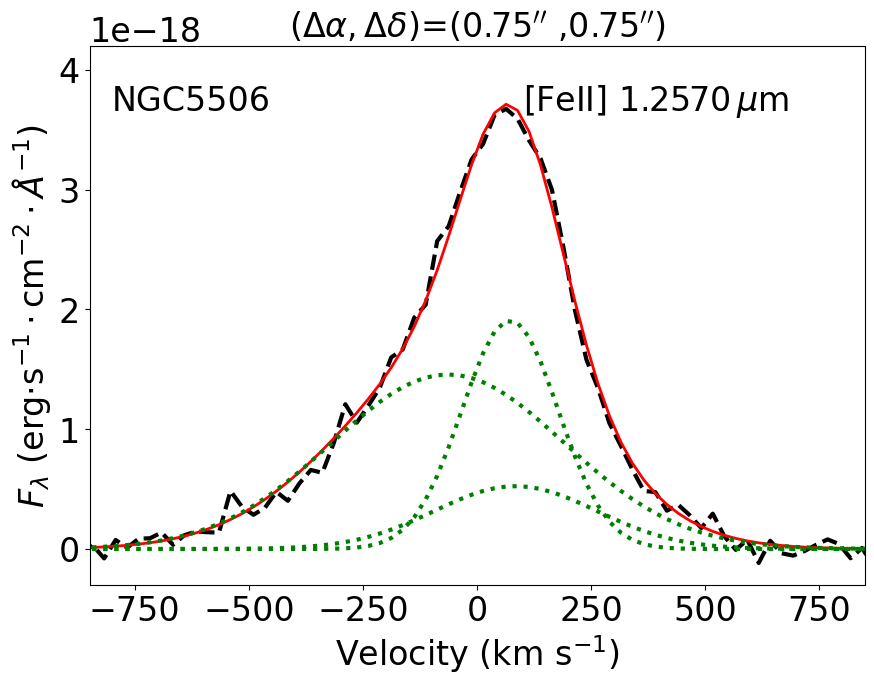} 
   \includegraphics[width=0.23\textwidth]{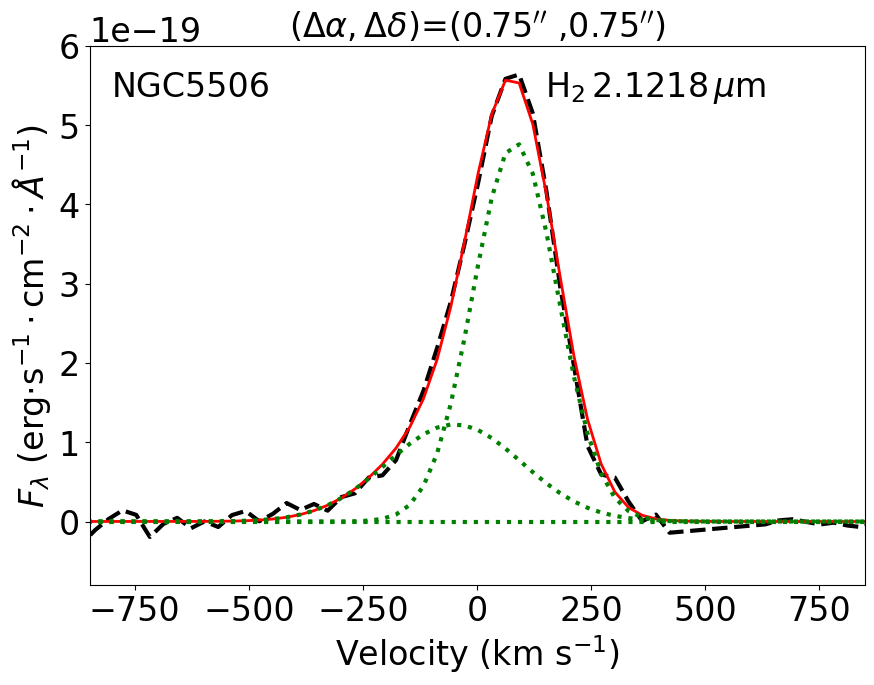}
\caption{Examples of fits of the \feiil\ (left) and \hml\ (right) emission-line profiles. The continuum-subtracted observed profiles are shown as dashed black lines, the model as continuous red lines and the individual Gaussian components as dotted green lines. The galaxy name is identified in the top-left corner  and the spaxel coordinates are shown in the title of each panel. } 
    \label{fig:fits}
\end{figure}

To investigate the origin of hot molecular and ionised gas emission in the galaxies of our sample, we measure the emission-line flux distribution by adopting the following procedure: first, we follow \citet{liu13} and  fit each emission line by a combination of three Gaussian functions using the {\sc ifscube} code \citep{ifscube}. The choice of the number of components has no physical motivation and simply aims to reproduce the observed profiles.
In previous papers by our group we used the {\sc profit} code \citep{rogemar_profit} to fit the emission-line profiles by Gaussian or Gauss-Hemite profiles \citep[e.g.][]{astor19}. The {\sc ifscube} code is more suitable for this work as it perform the fits of the emission-lines in each band simultaneously and allow the inclusion of more than one Gaussian component or Gauss-Hermite series. The underlying continuum is represented by a third order polynomial function.
Initial guesses for the centroid velocity and velocity dispersion of each component are provided to the code as obtained from the fit of the nuclear spaxel using the {\sc iraf.splot} task. These guesses are used by the code to model the nuclear spaxel and then  it performs the fitting of the neighboring spaxles following a spiral loop and using the parameters from spaxels located at distances smaller than 0.25\,arcsec from the fitted spaxel.   Although we allow up to three Gaussian components to fit each line profile,   if fewer components are able to properly reproduce the observed profile, the code fits only the number of components necessary to reproduce the observed profile, by setting the  amplitudes of the unnecessary  Gaussian functions as zero. Three galaxies of our sample (NGC\,3227, NGC\,3516, and NGC\,5506) host type 1 AGN. To fit the Pa$\beta$ and Br$\gamma$ line profiles of these galaxies, we included an additional Gaussian to account for the Broad Line Region (BLR) emission.  In Figure~\ref{fig:fits} we show examples of the fits of the [Fe\,{\sc ii}]$1.2570\,\mu$m (left) and H$_2\, 2.1218\,\mu$m (right) profiles for NGC\,788, NGC\,3227 and NGC\,5506.  

We integrate the fluxes in the modeled spectra within a spectral window of 1500 km\,s$^{-1}$ width centred at each emission line, after the subtraction of the contribution of the continuum emission. This procedure produces flux distributions very similar to those obtained from direct integration of the observed line profiles, but the modeled spectra is less sensitive to spurious features and thus the resulting flux maps from the modeled spectra are less affected by them. For the type 1 AGN, the BLR components are subtracted from the observed spectra before the computation of the fluxes of the narrow components.

\section{Results}
\label{sec:results}

\subsection{Emission-line flux distributions}

\begin{figure*}
    \centering
    \includegraphics[width=0.9\textwidth]{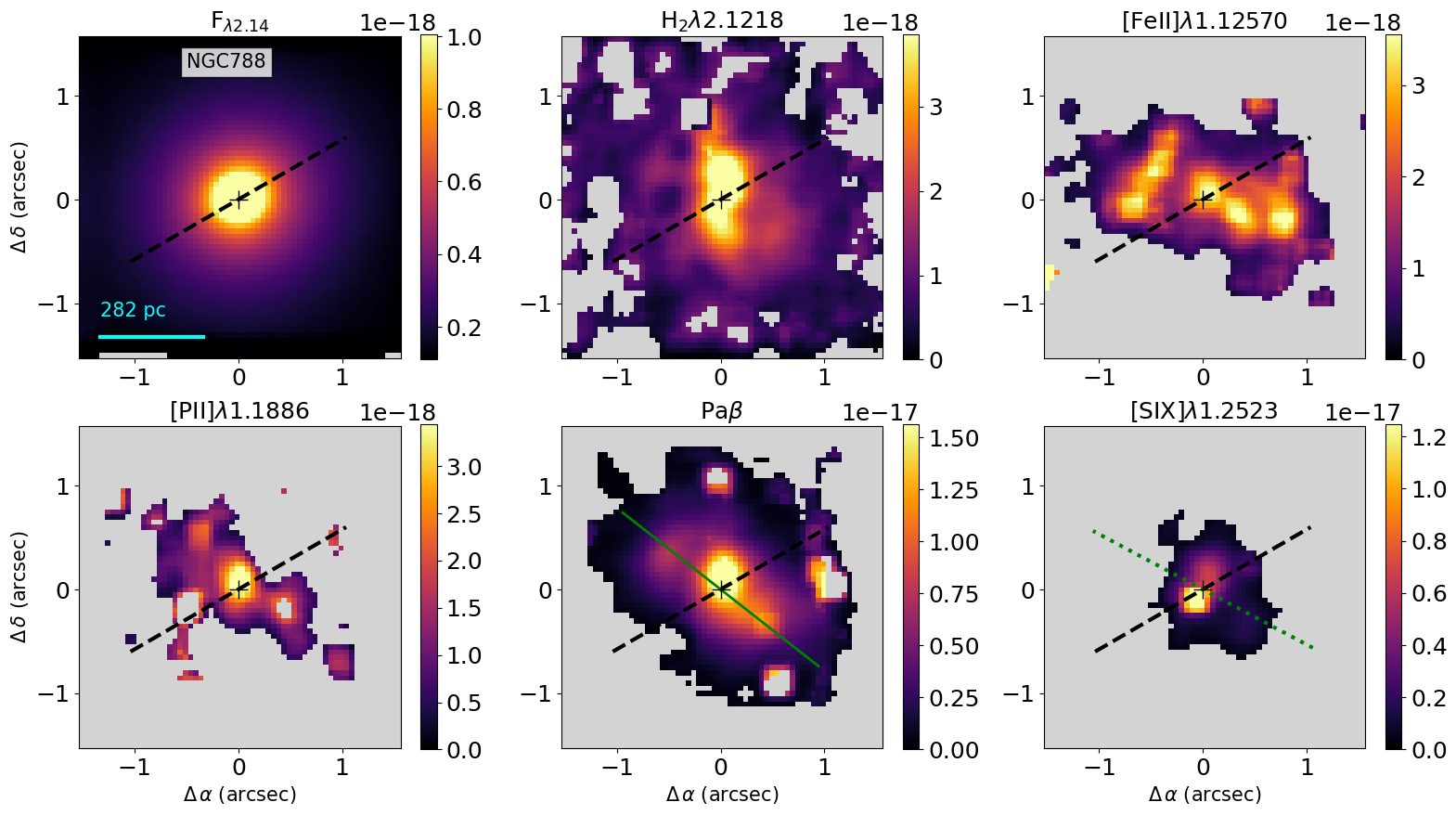}    
\includegraphics[width=0.9\textwidth]{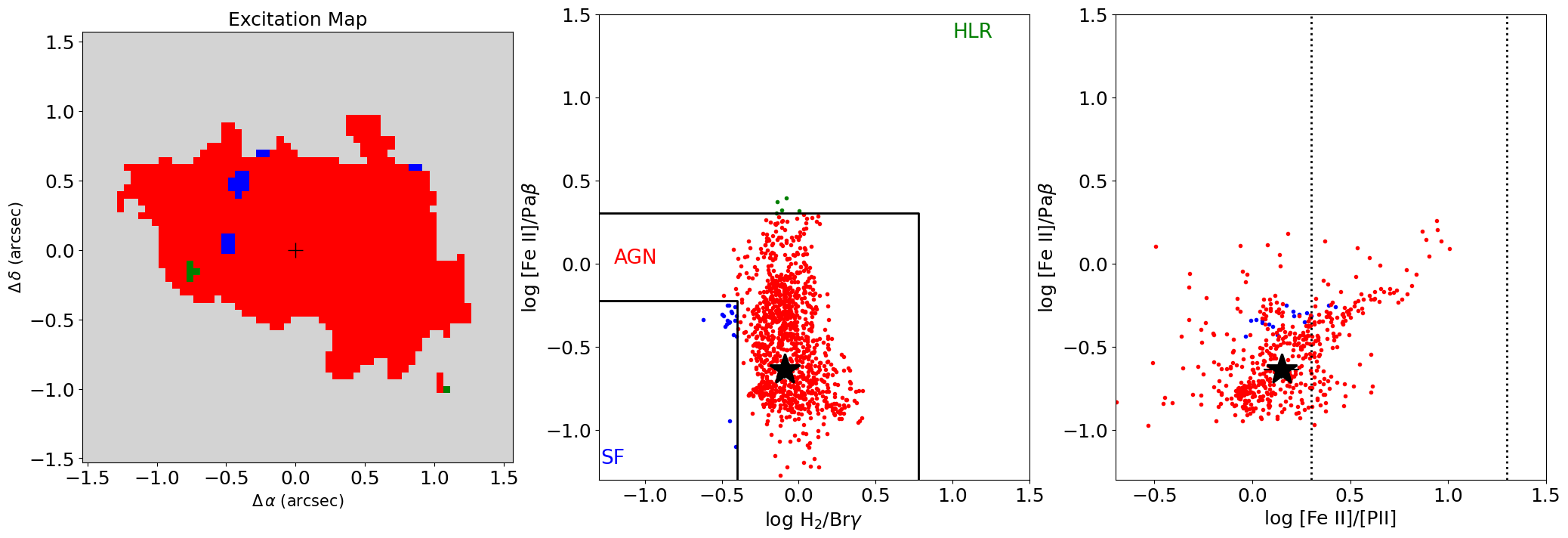}
   \includegraphics[width=0.9\textwidth]{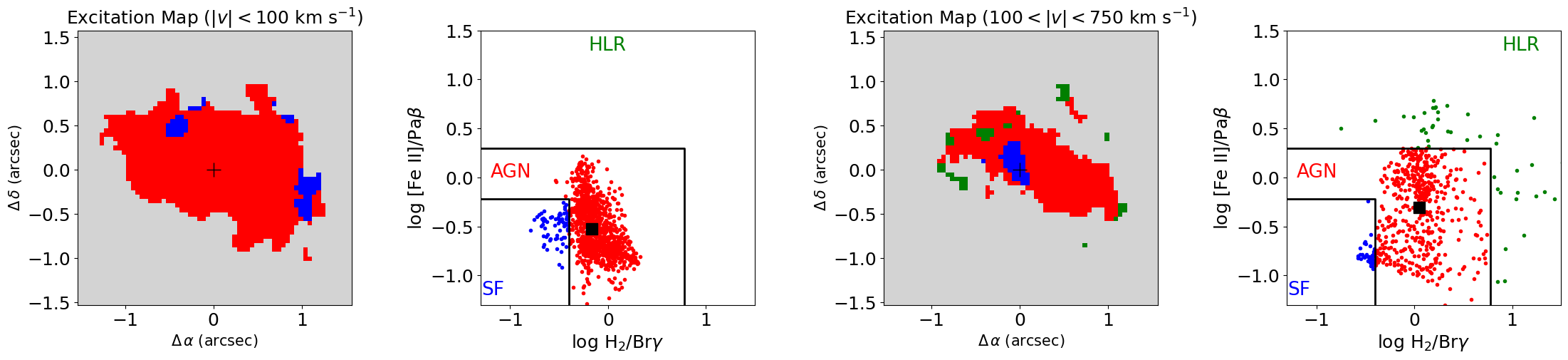} 
 \caption{{\bf NGC\,788} -- {\it 1st and 2nd rows}: Continuum emission measured from a K-band window and emission-line flux distributions for NGC\,788. Gray regions correspond to locations where the emission lines are not detected at a 3$\sigma$ continuum level, the crosses mark the location of the peak of the continuum emission and the dashed line corresponds to the orientation of the line of nodes derived by \citet{rogemar_stellar}. The green continuous line shown in the Pa$\beta$ map represents the orientation of the most collimated Pa$\beta$ emission (which can be used as an indicator of the AGN ionisation axis), computed using the {\sc cv2.moments} python package. The green dotted line overploted on the [S\,{\sc ix}] image shows the orientation of radio emission from \citet{nagar99}. The colour bars show the continuum in units of ${\rm erg s^{-1} cm^{-2}}$\AA$^{-1}$ and the line fluxes in units of ${\rm erg s^{-1} cm^{-2}}$ of each spaxel. North is up and east is to the left. 
 {\it 3rd row}: Emission-line ratio diagnostic diagrams for NGC\,788. The left panels show the excitation maps of the galaxy colour-coded according to the region in the \feiil/\pab\ vs. \hml/\brg\ diagram, shown in the central panels. The lines delineating the SF, AGN and high line ratio (HLR) regions are from \citet{rogerio13}. The right panels show the \feiil/\pab\ vs. \feiil/\piil\ diagram using the same colour scheme from the excitation map. The central dotted vertical line corresponds to typical line ratios from photoionised (\feii/[P\,{\sc ii}]$=$2), while the right dotted vertical line corresponds to shock-dominated  (\feii/[P\,{\sc ii}]$=$20) objects \citep{oliva01,sbN4151Exc}. The black stars show the line ratios within a circular aperture with radius of 0.5\,arcsec. {\it 4rh row}: Excitation maps and \feiil/\pab\ vs. \hml/\brg\ diagnostic diagrams for the low velocity ($|v|<100\,{\rm km\,s^{-1}}$: 1st and 2nd panels) and high velocity ($100\,{\rm km\,s^{-1}}<|v|<750\,{\rm km\,s^{-1}}$: 3rd and 4th panels) gas. The squares show the median line ratios.}
    \label{fig:N788}
\end{figure*}

\begin{figure*}
    \centering
    \includegraphics[width=0.9\textwidth]{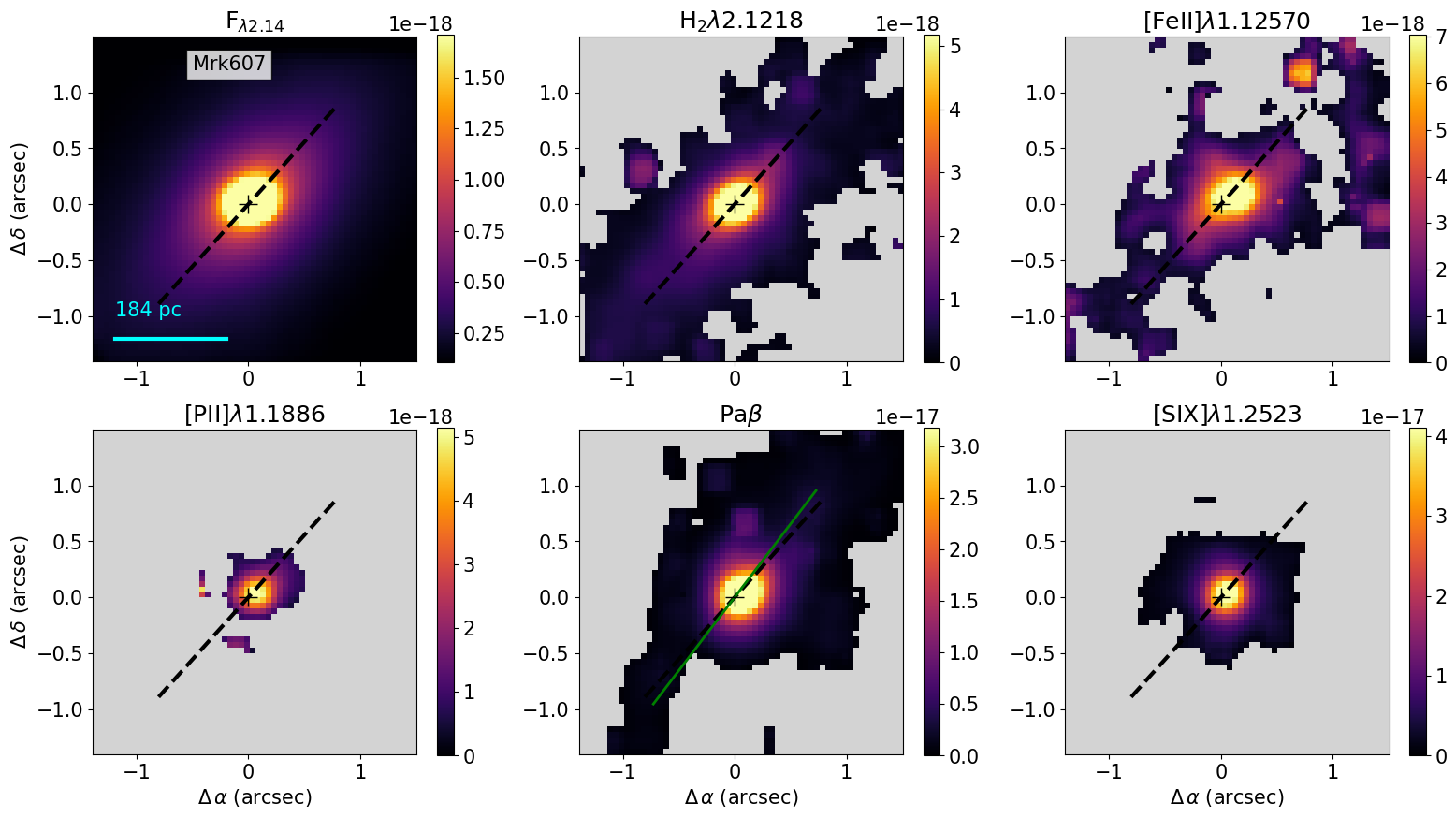}    
\includegraphics[width=0.9\textwidth]{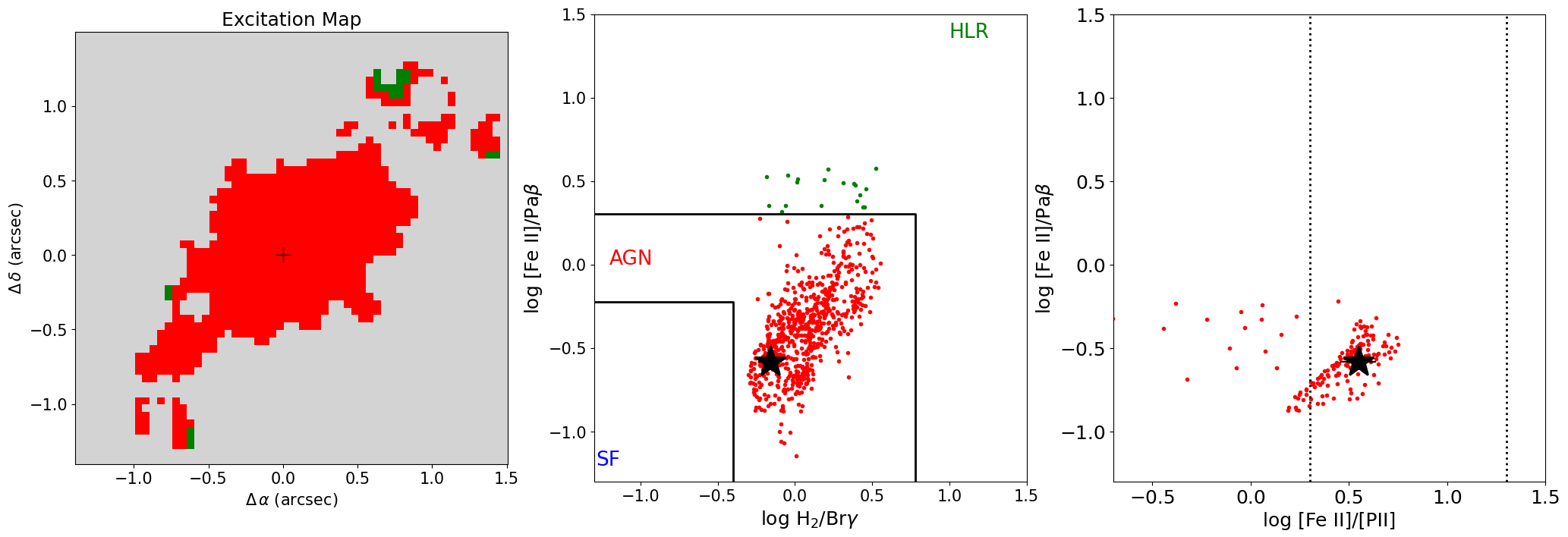}
   \includegraphics[width=0.9\textwidth]{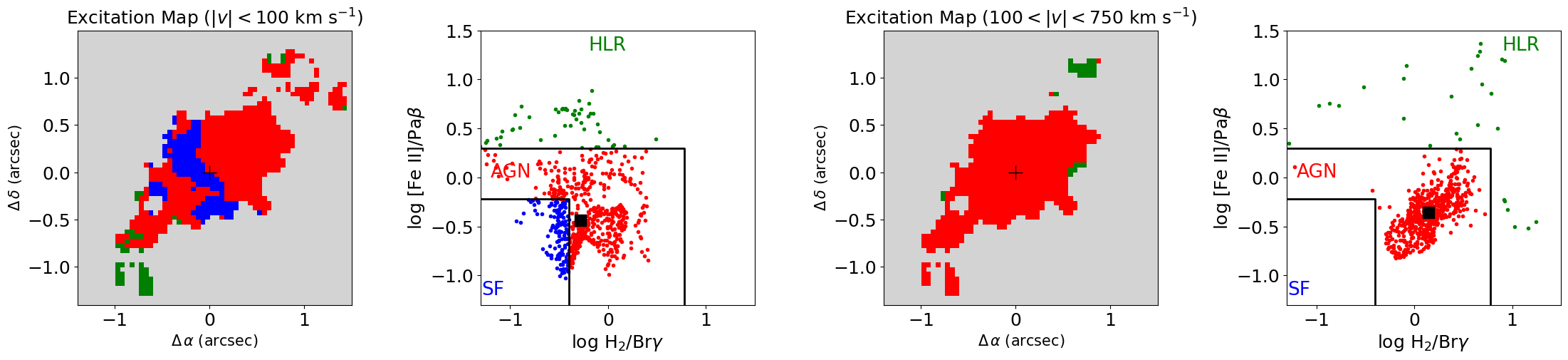} 
 \caption{Same as Fig.~\ref{fig:N788}, but for {\bf Mrk\,607}.}
    \label{fig:M607}
\end{figure*}

\begin{figure*}
    \centering
    \includegraphics[width=0.9\textwidth]{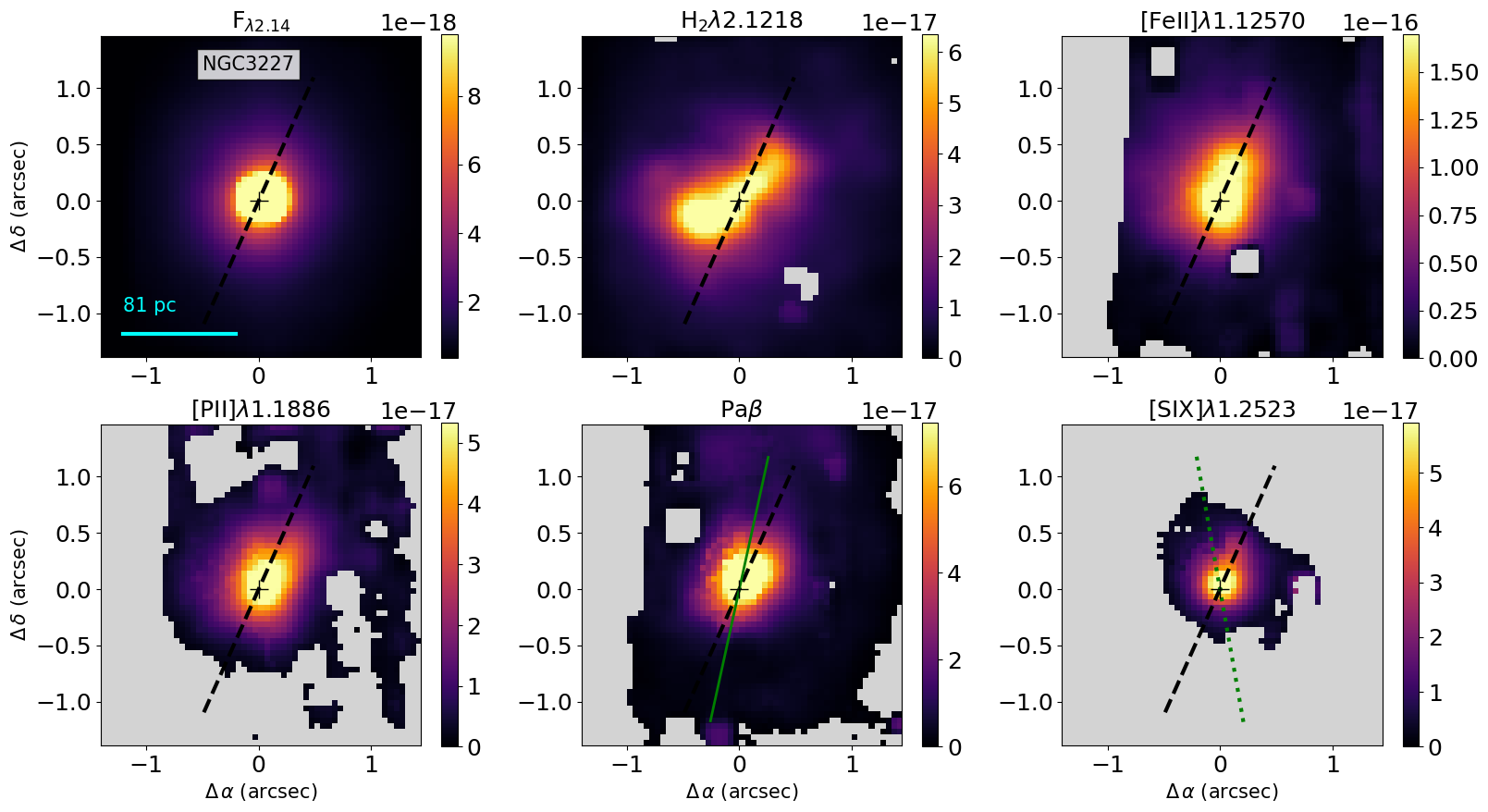}    
\includegraphics[width=0.9\textwidth]{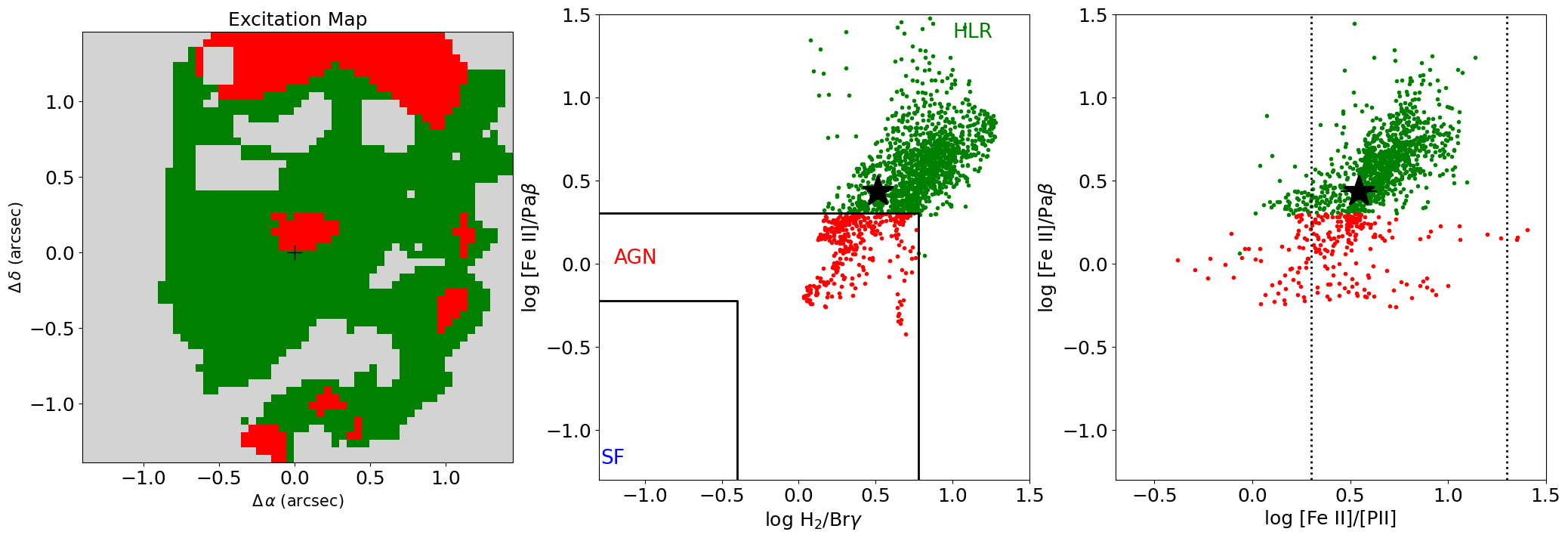}
   \includegraphics[width=0.9\textwidth]{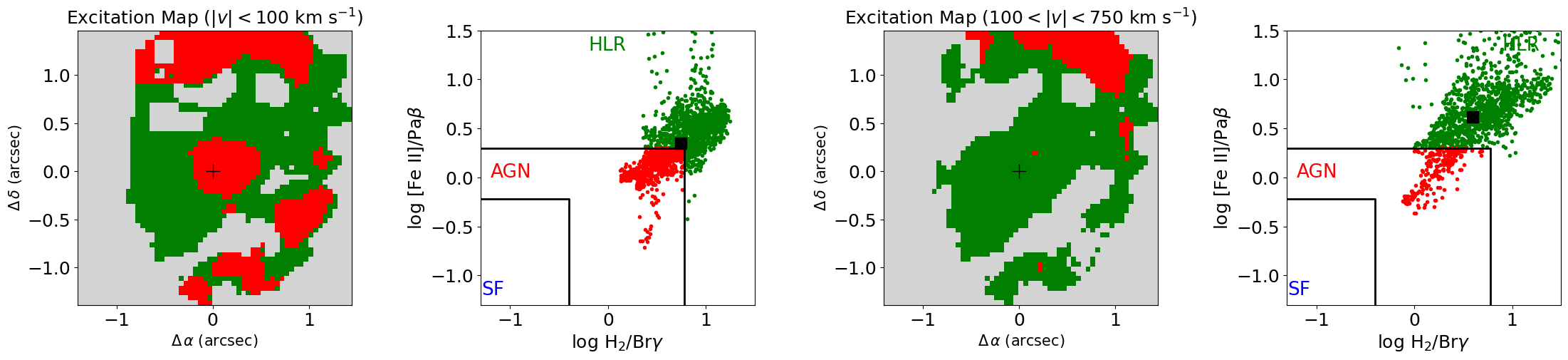} 
 \caption{Same as Fig.~\ref{fig:N788}, but for {\bf NGC\,3227}. The green dotted line overploted on the [S\,{\sc ix}] image shows the orientation of radio emission from \citet{mundell95}.}
    \label{fig:N3227}
\end{figure*}

\begin{figure*}
    \centering
    \includegraphics[width=0.9\textwidth]{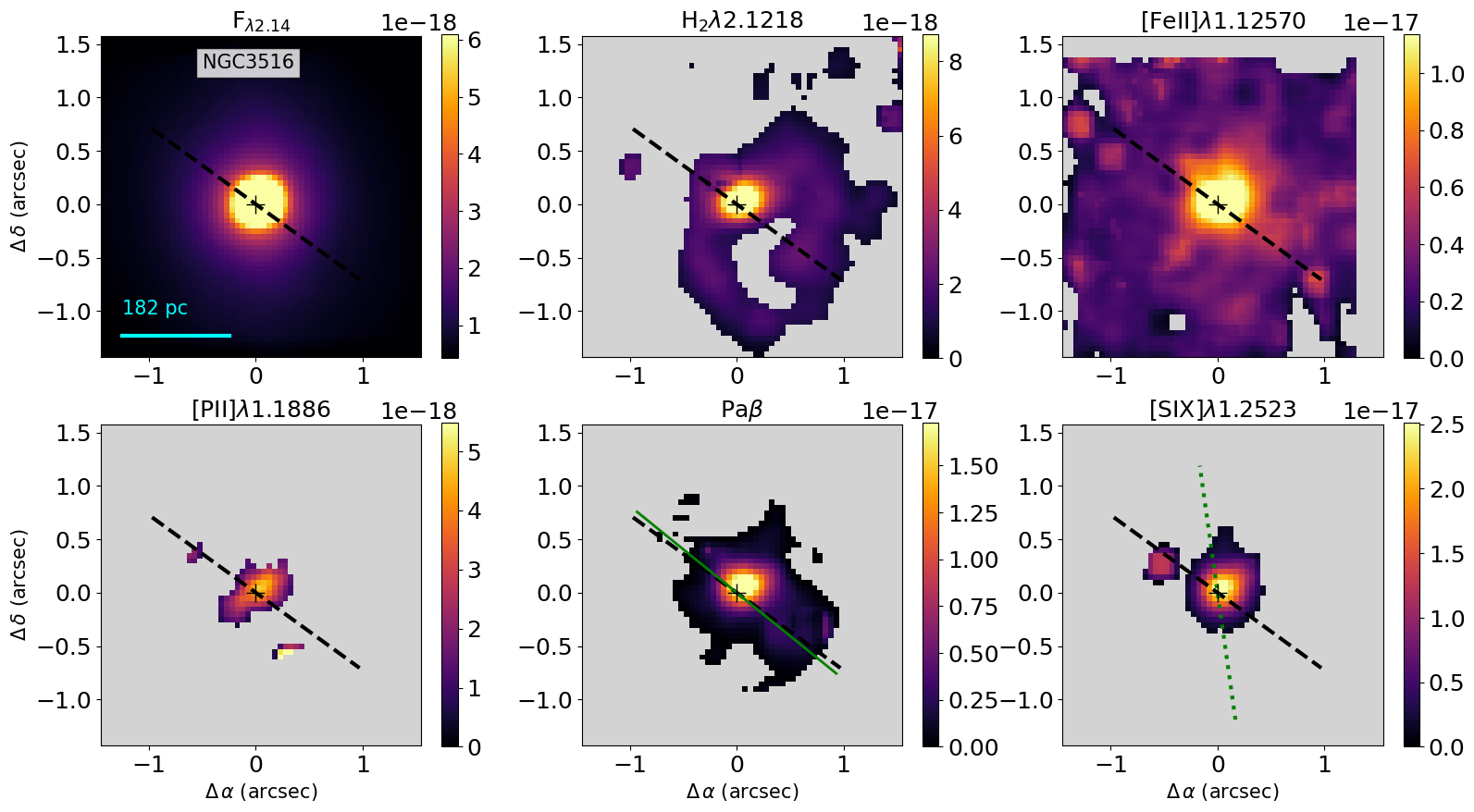}    
\includegraphics[width=0.9\textwidth]{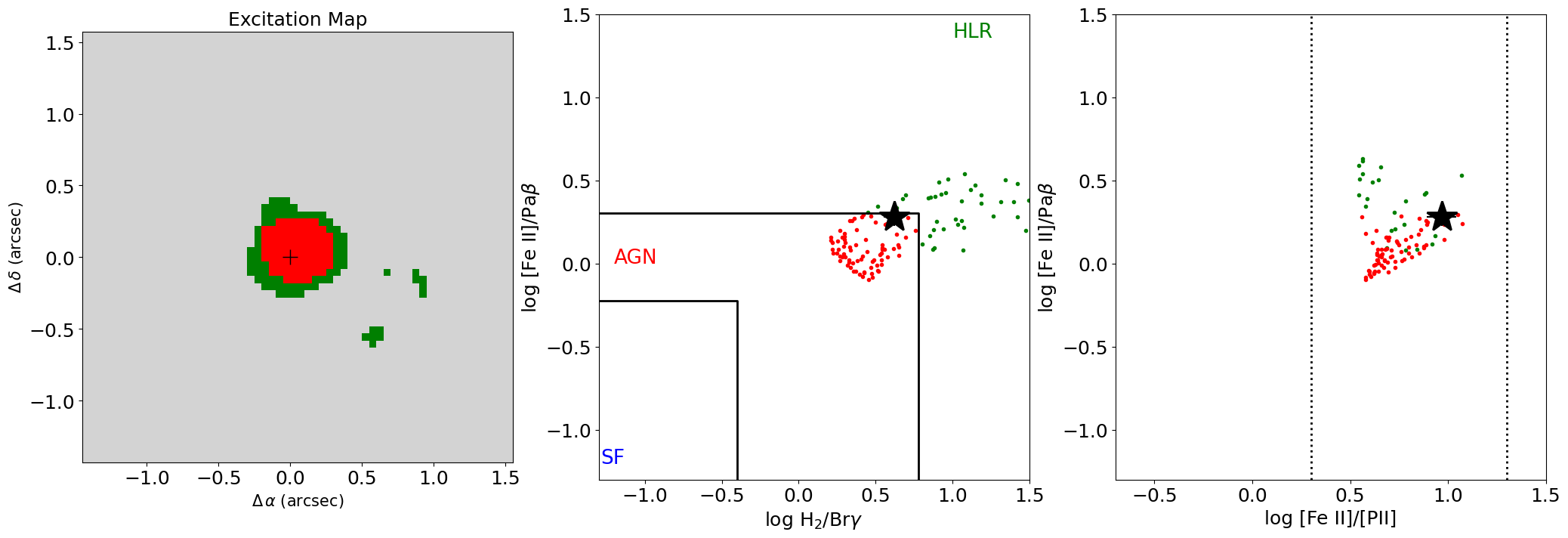}
   \includegraphics[width=0.9\textwidth]{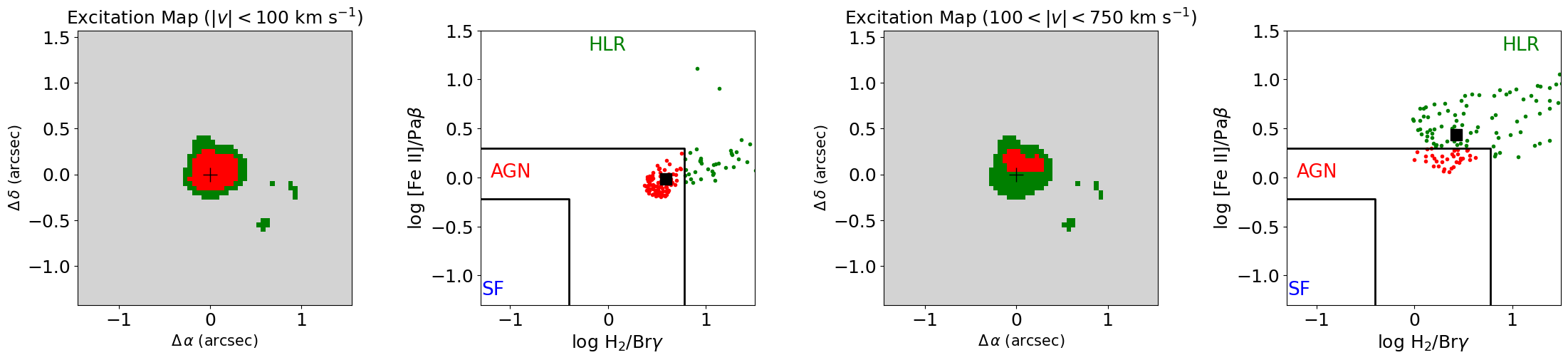} 
 \caption{Same as Fig.~\ref{fig:N788}, but for {\bf NGC\,3516}. The green dotted line overploted on the [S\,{\sc ix}] image shows the orientation of radio emission from \citet{nagar99}.} 
    \label{fig:N3516}
\end{figure*}

\begin{figure*}
    \centering
    \includegraphics[width=0.9\textwidth]{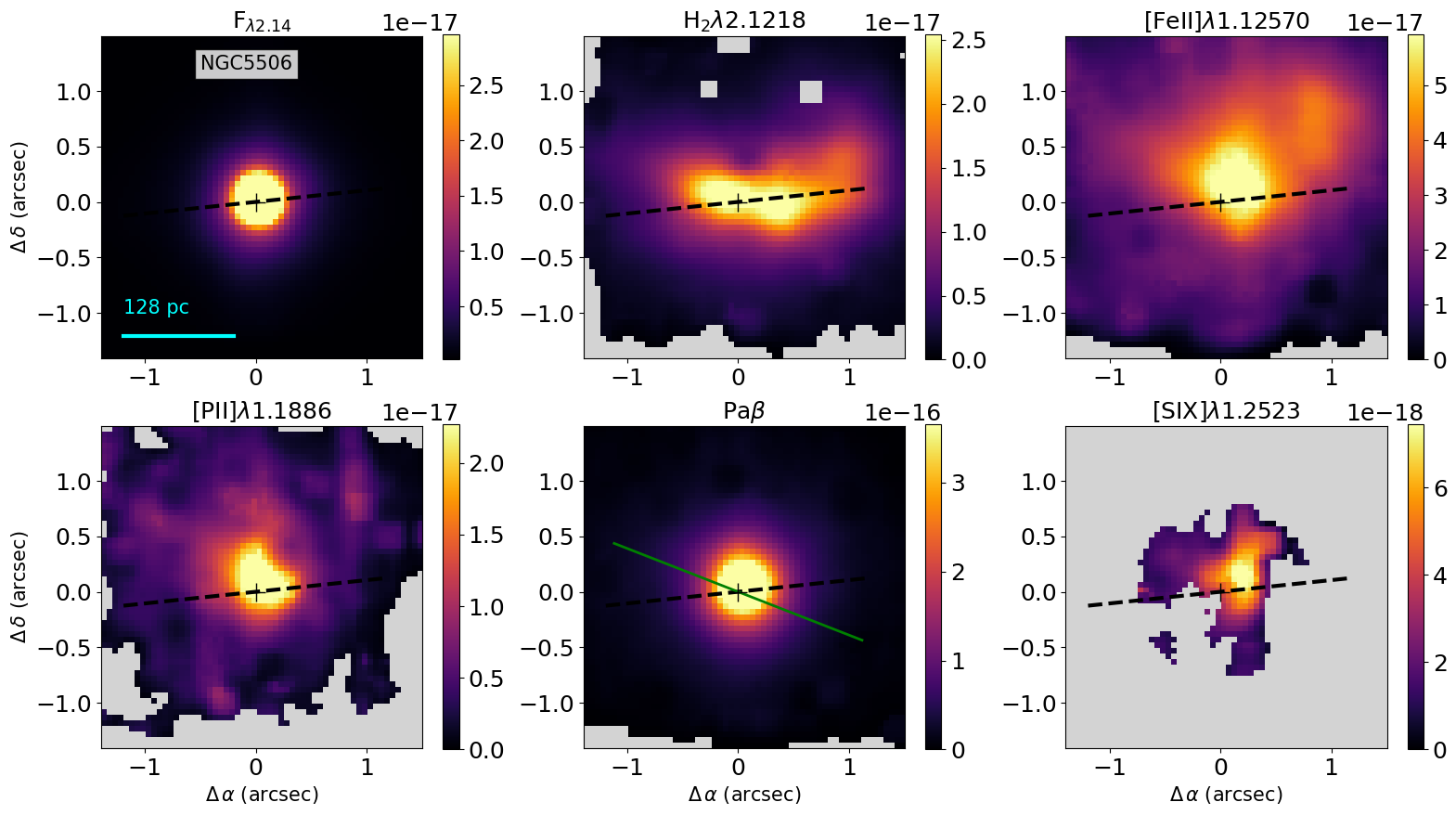}    
\includegraphics[width=0.9\textwidth]{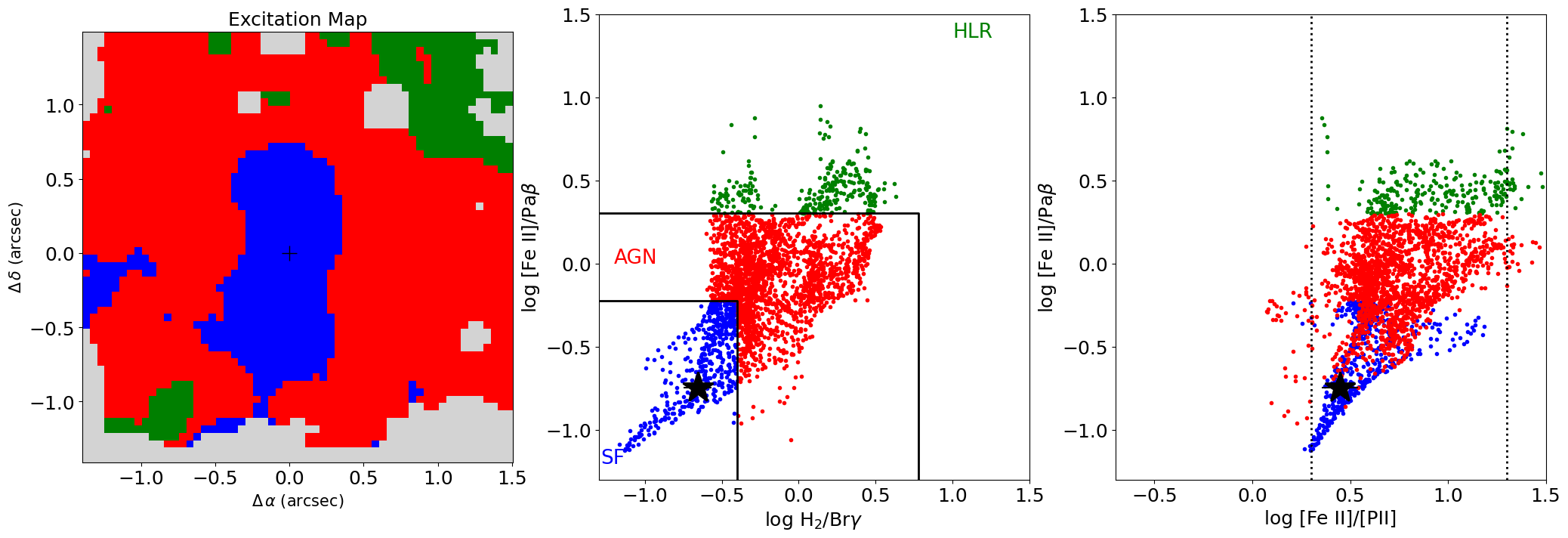}
   \includegraphics[width=0.9\textwidth]{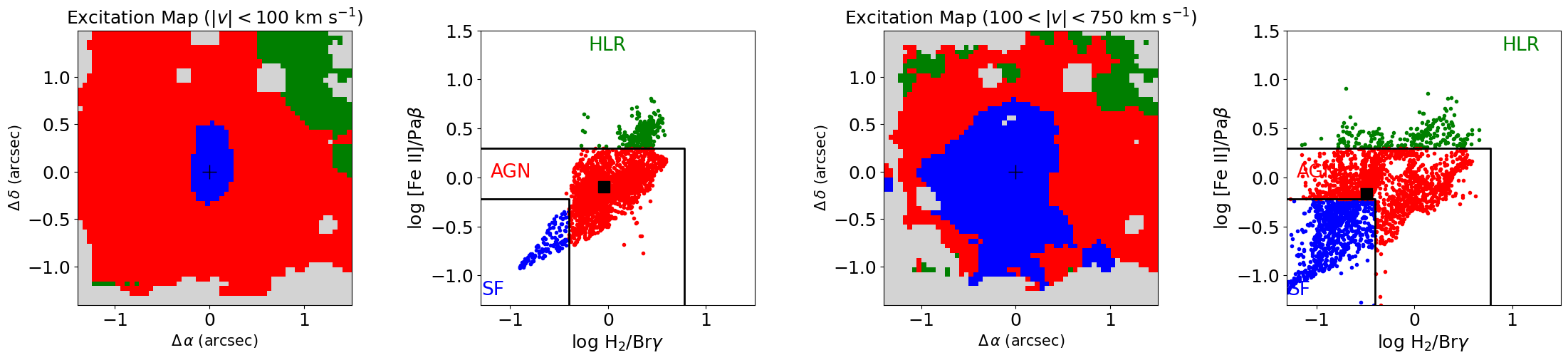} 
 \caption{Same as Fig.~\ref{fig:N788}, but for {\bf NGC\,5506}.}
    \label{fig:N5506}
\end{figure*}

\begin{figure*}
    \centering
    \includegraphics[width=0.9\textwidth]{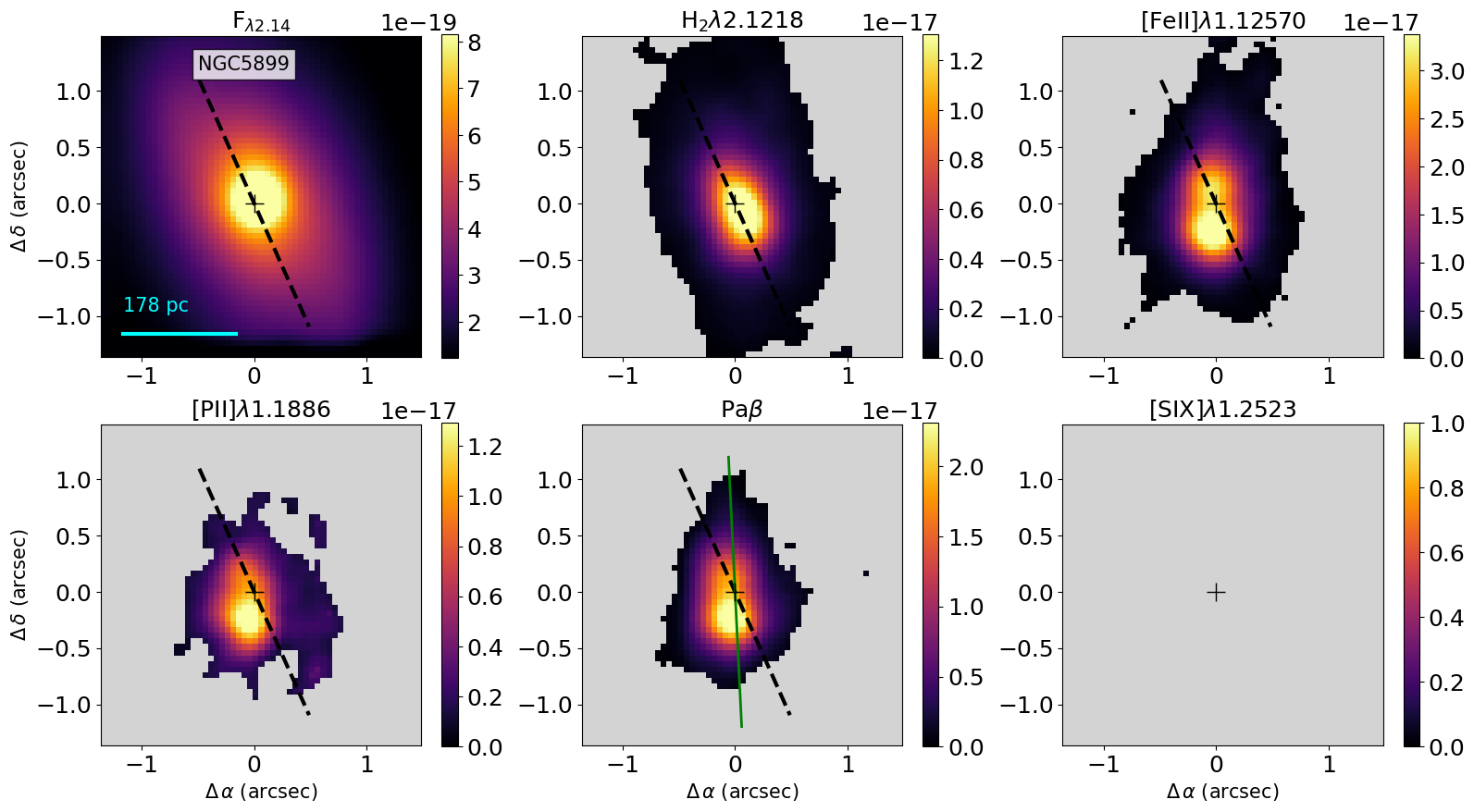}    
\includegraphics[width=0.9\textwidth]{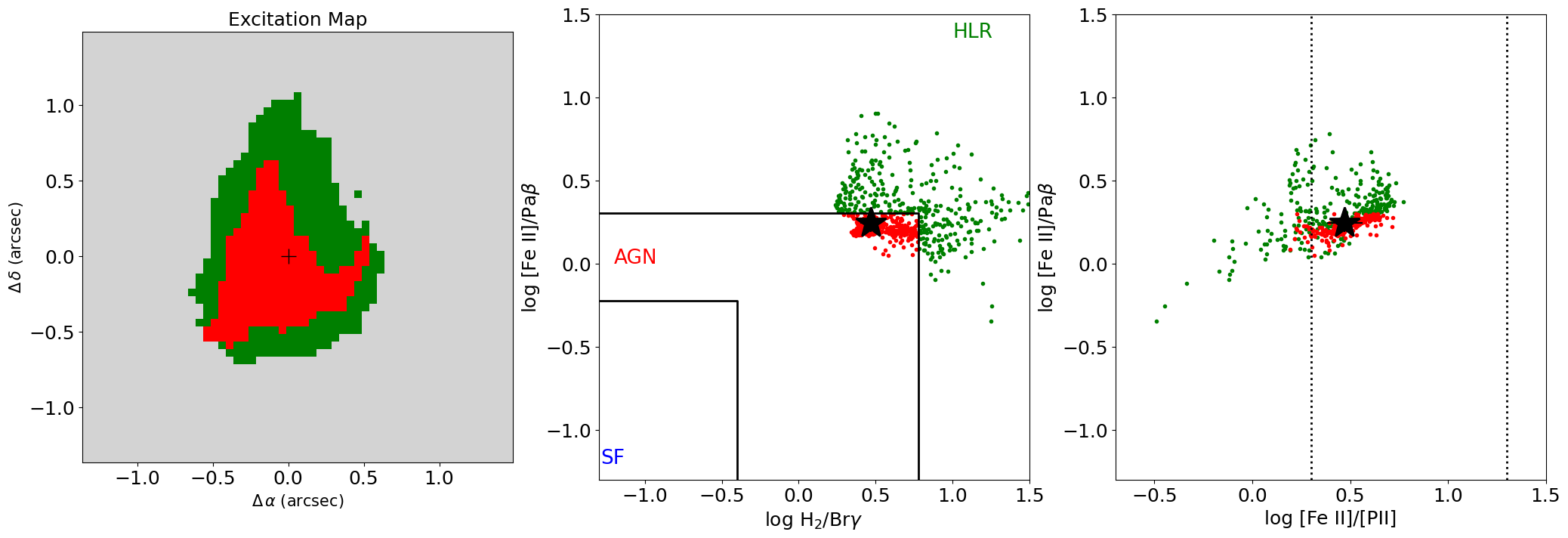}
   \includegraphics[width=0.9\textwidth]{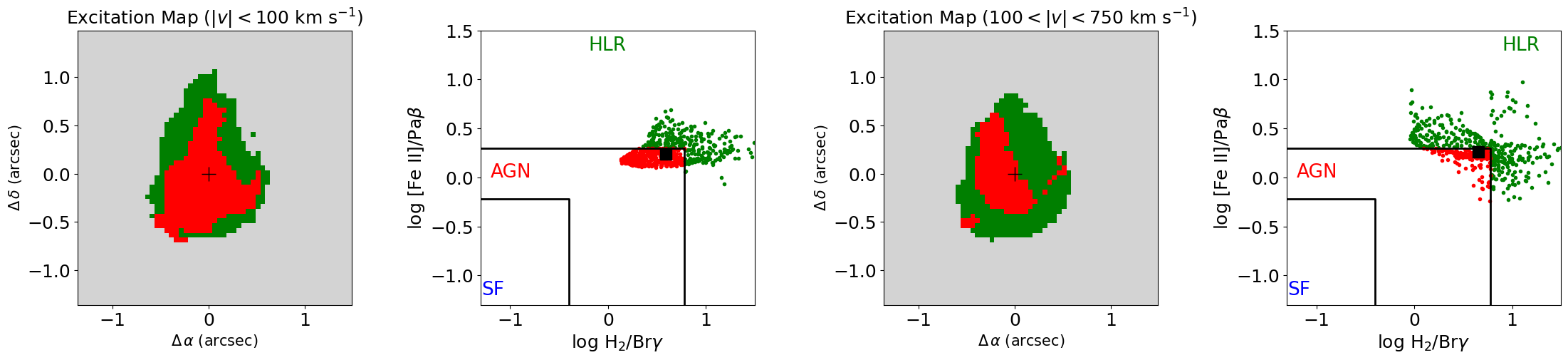} 
 \caption{Same as Fig.~\ref{fig:N788}, but for {\bf NGC\,5899}.}
    \label{fig:N5899}
\end{figure*}

The first and second rows of Figures~\ref{fig:N788} -- \ref{fig:N5899} 
present the K-band continuum image, \hml, \feiil, \piil, \pab\ and \sixl\ emission-line flux distributions of the galaxies of our sample. Although the H$_2$, \feii\ and \pab\ flux distributions were already shown in \citet{astor19},
the measurement method is distinct here, and the figures allow a comparison among the distinct emission line flux distribution for the different galaxies.
%In the case of NGC\,5506, instead of Pa$\beta$ the Br$\gamma$ is presented, as both emission lines trace the same gas. Our maps for these lines are consistent with those in \citet{astor19}, but we show them again to allow a better comparison among the distinct emission lines and galaxies.
In all maps, gray regions correspond to masked locations where the line is not detected at 3$\sigma$ level of the continuum noise. The dashed lines overlaid to all maps correspond to the orientation of line of nodes of the galaxy, as derived from the fitting of the stellar velocity fields by \citet{rogemar_stellar}.  The green continuous line shown in the Pa$\beta$ map shows the orientation of the most collimated Pa$\beta$ emission (which can be used as an indicator of the AGN ionisation axis), computed using the {\sc cv2.moments} python package. We performed Monte Carlo simulations with 100 iterations each, by adding random noise with amplitude of the 20th percentile flux value of the corresponding Pa$\beta$ map and use the average value of the position angle of all simulations. 
%We warn that this estimate is very dependent on the size of the FoV and consequently quite uncertain for the small NIFS FoV.

As already noticed by \citet{astor19}, distinct flux distributions are observed for the molecular and low ionised gas emission lines in most galaxies. The only exception is Mrk\,607, for which both molecular and low ionisation gas emission is observed mainly along the galaxy's major axis. We detect the \sixl\ emission in five galaxies of our sample. 
%This line has an ionisation potential of 328.2 eV and traces the emission of highly ionised gas, whose origin in AGN is still not well understood. 
The [S\,{\sc ix}] flux distribution extends to distances $\gtrsim$100 pc from the nucleus for  NGC\,788, Mrk\,607, NGC\,3227 and NGC\,5506, is marginally resolved for NGC\,3516 and is not detected in individual spaxels for NGC\,5899.

\subsection{Emission-line ratio diagnostic diagrams}

A way to investigate the excitation mechanisms of the near-IR emission lines is by constructing line-ratio diagrams \citep[e.g][]{reunanen02,ardila04,ardila05,rogerio13,rogemar_N1275}. The \feiil/\pab\ vs. \hml/\brg\ is the most commonly used diagram for this purpose. We present this diagram for the individual spaxels of the galaxies of our sample in the third row of Figures~\ref{fig:N788} -- \ref{fig:N5899}. The continuous lines show the empirical limits that separate Star Forming (SF) galaxies, AGN and high line ratio (HLR) objects as defined by \citet{rogerio13} using the nuclear spectra of a large sample of galaxies. The HLR region corresponds to the highest values of  \feiil/\pab\ and \hml/\brg\  line ratios, occupied by Transition Objects, LINERs and Supernovae Remnants. The H$_2$ flux maps show also emission from locations where no ionized gas emission is seen the H$_2$ emission is likely produced by shocks due to the interaction of a wide opening angle wind with the dense gas \citep[e.g.][]{zakamska14,rogemar_N1275} or by X-rays from the central AGN escaping through the dusty torus. The former is in agreement with shock models used to describe the H$_2$ molecule formation from luminous galaxies \citep{guillard09}. For the latter one would expect the regions with H$_2$ emission and no ionized gas emission being located mainly perpendicular to the AGN ionization axis, which is not supported by the flux distributions in our sample. Similarly, there is [Fe\,{\sc ii}] emission in regions outside the AGN ionization cones, likely tracing the photodissosiation regions at the cone edges. 

\citet{colina15} presented a detailed analysis of the two-dimensional gas excitation in a sample of LIRGs and Seyfert nuclei using IFS spectroscopy and defined new areas in the  near-IR diagnostic diagram for AGN, SF, and supernovae explosions. They used the [Fe\,{\sc ii}]1.6440\,$\mu$m/\brg\ and \hml/\brg\ line ratios, which can be converted to the ratios used in this work. \citet{colina15} used a similar range of values for the \hml/\brg\ to define the AGN region as that used in \citet{rogerio13}, but  their results indicate a larger upper limit of the \feiil/\pab\ (of $\sim$7, obtained using the relation [Fe\,{\sc ii}]1.6440\,$\mu$m/\brg$=4.4974\times$\feiil/\pab\ presented in their work).  The sample of Seyfert galaxies used in \citet{colina15} is composed of only 5 objects and all of them present outflows that can produce shocks %kinematic signatures of shocks
\citep{rogemar_eso428,rogemar_mrk1066_kin,rogemar_mrk1157,rogemar_mrk79,sb_N4151_kin,may18,may20}, and thus, a contribution of shocks to the \feii\ excitation cannot be ruled out. Thus, we used the separating lines from \citet{rogerio13},  that even if they are based on single aperture spectra, these authors used a larger sample, composed of 67 galaxies.

The left panels in the third row of  Figs.~\ref{fig:N788} -- \ref{fig:N5899} show the colour coded excitation maps, where SF dominated spaxels are shown in blue, AGN dominated are in red and spaxels with line ratios in the HLR region are in green.  All galaxies, but NGC\,5506, show nuclear line ratios typical of AGN. NGC\,788, Mrk\,607 and NGC\,3516 show line ratios in the AGN region of the diagnostic diagram in all locations, but the \brg\ emission line is detected only at the centre in the latter. HLRs are seen in some locations for NGC\,3227, NGC\,5506 and NGC\,5899. The only galaxy for which we found SF line ratios is NGC\,5506, observed in the innermost region around the nucleus. 

Another line ratio useful to investigate the role of shocks in the excitation of the \feii\ is the \feiil/\piil. This line ratio increases if shocks release the iron from dust grains reaching values of $\sim$20 in supernovae remnants, while values of $\lesssim$2 are seen in photoionised gas  \citep{oliva01,sbN4151Exc}. \citet{astor19} presented the \feii/\pii\ ratio maps for our sample and here we present a plot of \feii/\pab\ vs. \feii/\pii\ for the galaxy spaxels of our sample in the right panels of the third row of Figs.~~\ref{fig:N788} -- \ref{fig:N5899}. The colour coding is the same as for the  \feii/\pab\ vs. H$_2$/\brg\ diagram. The dotted vertical lines in these panels represent the limits of ratios in photoionised objects and in shock dominated objects -- \feii/[P\,{\sc ii}]$=2$ and 20, respectively \citep{oliva01,sbN4151Exc} -- see also \citet{rogerio19}. All galaxies show ratios larger than 2, with the highest values corresponding to the HLR regions in the excitation maps shown in the bottom-left panels of Figs.~~\ref{fig:N788} -- \ref{fig:N5899}. 

Table\,\ref{tab:ratios} presents the emission-line ratios, measured within two distinct apertures: (i) within the 100 pc radius and (ii) within 0.5\,arcsec radius. The first aperture allows the comparison of the line ratios observed in distinct galaxies at the same physical scale, while the latter is useful to investigate the nuclear line emission. 

The bottom row of Figs.~\ref{fig:N788} -- \ref{fig:N5899} show the  excitation maps and diagnostic diagrams for each galaxy, constructed using the line fluxes integrated within two velocity ranges: (i) the low velocity range $|v|<100\,{\rm km\,s^{-1}}$ and (ii) the high velocity range  $100\,{\rm km\,s^{-1}} < |v|<750\,{\rm km\,s^{-1}}$. The velocity range in each spaxel is measured relative to the velocity of the peak of the corresponding line profile. The high velocity range is more sensitive to shock excitation, as observed in optical emission lines \citep[e.g.][]{ho14,dagostino19b,rogemar20_letter}.  We exclude from these diagrams spaxels where the emission lines are not detected at a 3$\sigma$ continuum level.
%where the line fluxes in the considered velocity range are smaller than 1\,\% of the integrated line flux.  The black squares show the median values of each diagram.   

%#PA radio
%NGC788  62   https://ui.adsabs.harvard.edu/abs/1999ApJS..120..209N/abstract
%Mrk607  0?   https://ui.adsabs.harvard.edu/abs/1999ApJS..120..209N/abstract
%NGC3516 0 *to the north*   https://ui.adsabs.harvard.edu/abs/1999ApJS..120..209N/abstract
%NGC3227 18.5                http://articles.adsabs.harvard.edu/pdf/1995MNRAS.276.1262K
%NGC5506 70  https://ui.adsabs.harvard.edu/abs/2001ApJS..132..199S/abstract
%NGC5899  ND

\begin{table*}
	\centering
	\caption{\feiil/\pab, \feiil/\piil, \sixl/\pab\ and \hml/\brg\ emission-line ratios computed within circular apertures of 100 pc and  0.5\,arcsec radii.}
	\label{tab:ratios}
	\begin{tabular}{lcccccccc} % four columns, alignment for each
\hline
 & \multicolumn{4}{c}{aperture of 100 pc radius} & \multicolumn{4}{c}{aperture of 0.5\,arcsec radius}\\
Galaxy & [Fe\,{\sc ii}]/Pa$\beta$ & [Fe\,{\sc ii}]/[P\,{\sc ii}] & [S\,{\sc ix}]/Pa$\beta$ & H$_2$/Br$\gamma$ &  [Fe\,{\sc ii}]/Pa$\beta$ & [Fe\,{\sc ii}]/[P\,{\sc ii}] &  [S\,{\sc ix}]/Pa$\beta$& H$_2$/Br$\gamma$ \\
\hline
NGC788 & 0.18$\pm$0.05 & 1.15$\pm$0.55 & 0.39$\pm$0.16 & 0.71$\pm$0.25 & 0.23$\pm$0.03 & 1.41$\pm$0.33 & 0.32$\pm$0.07& 0.81$\pm$0.14 \\
Mrk607 & 0.26$\pm$0.03 & 2.98$\pm$0.57 & 0.75$\pm$0.13 & 0.70$\pm$0.11 &0.26$\pm$0.03 & 3.53$\pm$0.82 & 0.76$\pm$0.17 & 0.69$\pm$0.13\\
NGC3227 & 3.44$\pm$0.05 & 5.12$\pm$0.15 & 0.66$\pm$0.11  & 5.12$\pm$0.27 &  2.71$\pm$0.35 & 3.5$\pm$0.82 & 0.42$\pm$0.09 & 1.44$\pm$0.25\\ 
NGC3516 & 1.70$\pm$0.24 & 8.50$\pm$1.98 & 1.23$\pm$0.25  &  4.40$\pm$0.64 & 1.9$\pm$0.24 & 9.28$\pm$1.79 & 1.26$\pm$0.22 & 4.19$\pm$0.73 \\
NGC5506 & 0.25$\pm$0.05 & 3.46$\pm$0.32 & 0.011$\pm$0.002  & 0.28$\pm$0.05 &  0.18$\pm$0.02 & 2.81$\pm$0.66 & 0.011$\pm$0.004 & 0.22$\pm$0.05\\ 
NGC5899 &     1.72$\pm$0.18 & 3.07$\pm$0.59 & 0.001$\pm$0.001  & 3.03$\pm$0.25 &  1.75$\pm$0.23 & 2.97$\pm$0.69 & 0.001$\pm$0.001 & 2.94$\pm$0.51\\
	\hline
	\end{tabular}
\end{table*}
\begin{figure}
    \centering
\includegraphics[width=0.35\textwidth]{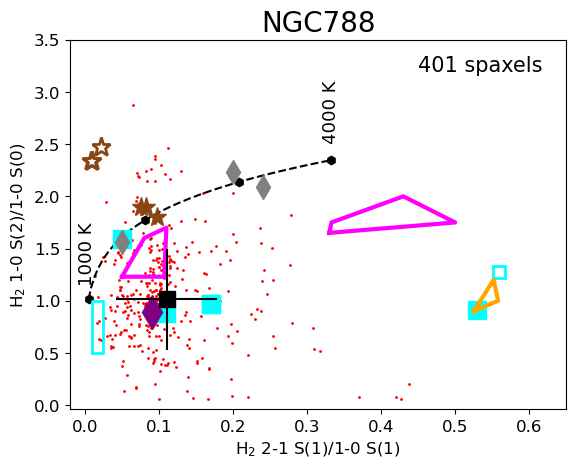} 
\includegraphics[width=0.35\textwidth]{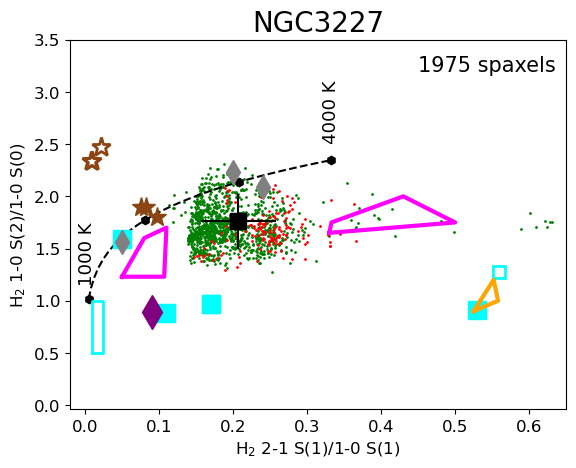} 
\includegraphics[width=0.35\textwidth]{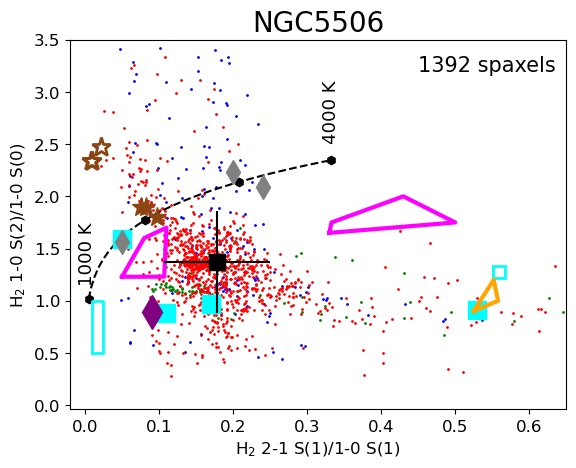} 
\caption{H$_2$\,2--1\,S(1)$2.2477\,\mu$m/1--0\,S(1)$2.1218\,\mu$m vs. 1--0\,S(2)$2.0338\,\mu$m/1--0\,S(0)$2.2235\,\mu$m diagnostic diagram for NGC\,788, NGC\,3227 and NGC\,5506. The colours of the points represent the different regions identified in the diagnostic diagrams of Figs.~\ref{fig:N788}, \ref{fig:N3227} and \ref{fig:N5506}. The black dashed curve corresponds to the ratios for an isothermal and uniform density gas distribution for temperatures ranging from 1000 to 4000\,K, rom left to right. The open cyan rectangle shows the regions predicted from the thermal UV excitation models of \citet{Sternberg89} for gas densities ($n_t$) between 10$^5$ and 10$^6$ cm$^{-3}$ and UV scaling factors relative to the local interstellar radiation field $\chi$ from 10$^2$ to 10$^{4}$, as computed by \citet{mouri94}. The open cyan square is from \citet{Sternberg89} for $n_t=10^3$ cm$^{-3}$ and $\chi=10^2$ and the filled cyan squares are the UV thermal models from \citet{davies03} with $10^3<n_t<10^6$ cm$^{-3}$ and $10^2<\chi<10^5$.   The brown open and filled stars are from the thermal X-ray models of \citet{Lepp83} and \citet{draine90}, respectively. The purple diamond is from the shocks model of \citet{Kwan77} and the gray diamonds represent the shock models from \citet{Smith95}. The orange polygon represents the region occupied by the non-thermal UV excitation models of \citet{black87} and the open magenta polygons cover the region of the photoionisation models of \citet{dors12}.  We do not overlay the predictions of the photoionisation from \citet{rogerio13}, as they span a wide range in both axes ($0\lesssim$2--1 S(1)/1--0 S(1)$\lesssim0.6$  and  $0.5\lesssim$1--0 S(2)/1--0 S(0)$\lesssim2.5$). The black squares with error bars show the median line ratios and standard deviations for each galaxy.} 
    \label{fig:diagrams}
\end{figure}

The H$_2$\,2-1\,S(1)$2.2477\,\mu$m/1-0\,S(1)$2.1218\,\mu$m vs. 1-0\,S(2)$2.0338\,\mu$m/1-0\,S(0)$2.2235\,\mu$m  diagram is useful to investigate the origin of the H$_2$ emission. The spectral range for Mrk\,607 and NGC\,5899 does not include the H$_2\,2.0338\,\mu$m emission line and these lines are not detected in our data of NGC\,3516. Thus, we construct the H$_2$ diagnostic diagrams only for NGC\,788, NGC\,3227 and NGC\,5506 and present them in Figure~\ref{fig:diagrams}. The colours of the points are defined as in Figs.~\ref{fig:N788} -- \ref{fig:N5899} and we show also in the figure the predictions of distinct models.
%Although all galaxies present similar range of values of H$_2$ 2-1\,S(1)/1-0\,S(1), the 1-0\,S(2)/1-0\,S(0) ratio values are distinct:  for NGC\,788 they are $\lesssim1$, for NGC\,3227 the observed ratios are $\gtrsim1.5$ and NGC\,5506 presents values in the range $\sim$1.0--2.5.  
Most points in the diagrams for the three galaxies are away from the region predicted by the non-thermal UV excitation models of \citet{black87} (identified by the orange polygon), indicating that the H$_2$ emission originates in thermal processes.  
 
\subsection{Molecular hydrogen temperatures}

\begin{figure*}
\includegraphics[width=0.6\textwidth]{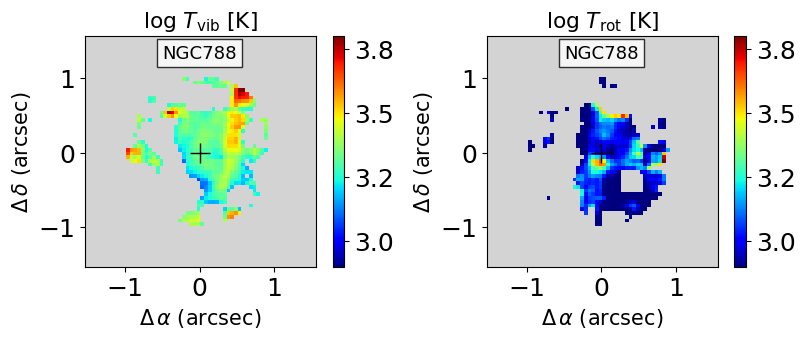} 
\hspace{1.0cm}
\includegraphics[width=0.3\textwidth]{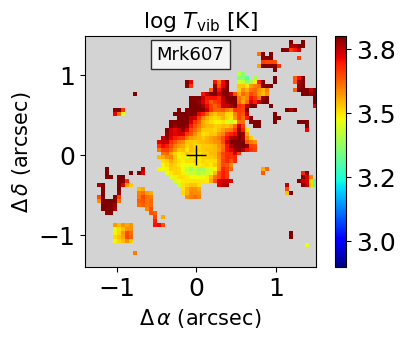} 
\includegraphics[width=0.6\textwidth]{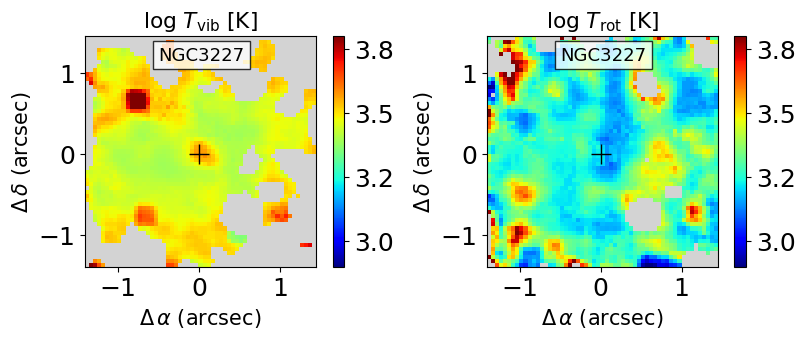} 
\hspace{1.0cm}
\includegraphics[width=0.3\textwidth]{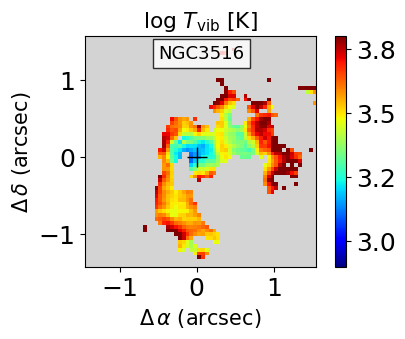}
\includegraphics[width=0.6\textwidth]{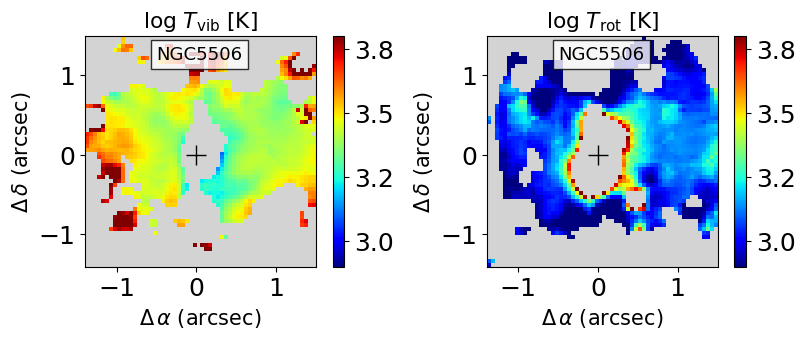} 
\hspace{1.0cm}
\includegraphics[width=0.3\textwidth]{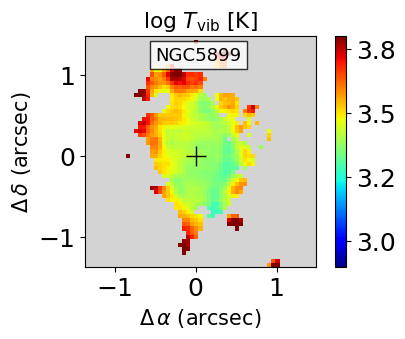}
\caption{H$_2$ vibrational and rotational temperatures of NGC\,788, NGC\,3227 and NGC\,5506 calculated according to Eqs.~\ref{eq:temp_vib} and \ref{eq:temp_rot}. For Mrk\,607, NGC\,3516 and NGC\,5899 only the vibrational temperature is presented because the H$_2 2.0338\,\mu$m line is not detected in these galaxies. The gray regions indicate the spaxels where at least one of the H$_2$ emission lines is not detected within 2$\sigma$ above the continuum noise. }
 \label{fig:temperarutes}
\end{figure*}

Figure~\ref{fig:temperarutes} presents the vibrational ($T_{\rm vib}$) and rotational ($T_{\rm rot}$) temperature distributions of the H$_2$ gas. These temperatures are obtained using the fluxes of the H$_2$ lines according to
\begin{equation}
T_{\rm vib}\cong {\frac{5600}{\ln\left(1.355\frac{F_{\rm H_{2}2.1218\mu m}}{F_{\rm H_{2}2.2477\mu m}}\right)}}.
\label{eq:temp_vib}
\end{equation}
and
\begin{equation}
T_{\rm rot} \cong -{\frac{ 1113}{\ln\left(0.323\frac{F_{\rm H_{2}2.0338\mu m}}{F_{\rm H_{2}2.2235\mu m}}\right)}}, 
\label{eq:temp_rot}
\end{equation}
respectively \citep{reunanen02} and the Einstein coefficients are taken from \citet{turner77}.  As mentioned above, for Mrk\,607 and NGC\,5899 the spectral range does not include the H$_2\,2.0338\,\mu$m emission line and thus we do not estimate the H$_2$ rotational temperatures for these galaxies. Although the spectral region of the NGC\,3516 data includes the H$_2 2.0338\,\mu$m emission line, the signal-to-noise ratios in both lines used to derive the rotational temperature are not high enough to obtain $T_{\rm rot}$, so we also present only the vibrational temperature map for this galaxy. The gray regions in the maps of Fig.~\ref{fig:temperarutes} correspond to locations where one or both emission lines used to calculate the corresponding temperature are not detected with, at least, 2$\sigma$ above the noise level with respect to the adjacent continuum.

The vibrational temperatures range from $\sim$1000 to $\sim$5000\,K, while the rotational temperatures range from a few hundred to $\sim$4000\,K. Overall, NGC\,788 shows much smaller values of $T_{\rm rot}$ than those of $T_{\rm vib}$, while for NGC\,3227 both temperature ranges are similar and for NGC\,5506 $T_{\rm rot}$ is slightly smaller than $T_{\rm vib}$.  Table~\ref{tab:temp} presents temperature values estimated within circular apertures of 100\,pc and 0.5\,arcsec radii, centred at the nuclei of the galaxies.

\begin{table*}
	\centering
	\caption{H$_2$ vibrational and rotational temperatures computed within circular apertures of 100 pc (columns 2 and 3) and 0.5\,arcsec (columns 4 and 5) radius. }
	\label{tab:temp}
	\begin{tabular}{lcccc} % four columns, alignment for each
\hline
Galaxy & $T_{\rm vib: r\leq100\,pc}$ (K) & $T_{\rm rot: r\leq100\,pc}$ (K) &$T_{\rm vib: r\leq 0.5^{\prime\prime}}$ (K) & $T_{\rm rot: r\leq0.5^{\prime\prime}}$ (K) \\
\hline
NGC788 & 2145$\pm$308 & 1126$\pm$963 & 2276$\pm$201 & 1073$\pm$339 \\
Mrk607 & 3634$\pm$419 & -- & 3370$\pm$422 & -- \\
NGC3227 & 3069$\pm$69 & 1745$\pm$130 & 2864$\pm$313 & 1723$\pm$820\\ 
NGC3516 & 2546$\pm$213 & -- & 2431$\pm$228 & --\\
NGC5506 & 2055$\pm$72 & 1485$\pm$242& 1602$\pm$102 & 1528$\pm$860\\ 
NGC5899 & 2438$\pm$196 & -- & 2525$\pm$245 & --\\
	\hline
	\end{tabular}
\end{table*}

\section{Discussion}
\label{sec:discussion}

\subsection{Notes on individual objects}

\subsubsection{NGC\,788} 
NGC\,788 is a spiral galaxy (SA0/a?(s)) located at 56\,Mpc and hosts a Sy\,2 nucleus. The stellar velocity field is described by a rotation pattern, with the orientation of the line of the nodes $\Psi_0\sim120^{\circ}$ and smaller ($\sim50-80$\,km\,s$^{-1}$) stellar velocity dispersion patches are observed at 250\,pc, probably due to intermediate age stellar populations \citep{rogemar_stellar}. The radial profile for the ionised gas mass is more concentrated in the centre  ($\sim 100$\,pc) and has a steeper distribution than that of the hot molecular gas \citep{rogemar_sample}. By mapping the gas distribution, \citet{astor19} show that Pa$\beta$ and \feii$\,1.2570\,\mu$m emission are extended along the northeast--southwest direction and present higher ($\sim 150$\,km\,s$^{-1}$) velocity amplitudes and regions of $\sigma\sim150$\,km\,s$^{-1}$\ in the same direction, interpreted as a bipolar outflow. On the other hand, the H$_2\,2.1218\,\mu$m presents emission in the whole NIFS FoV and low velocity dispersion ($\sigma\sim50$\,km\,s$^{-1}$).  NGC\,788 presents a compact radio emission at 3.6\,cm and slightly resolved extended radio emission along the position angle $\rm PA=62^\circ$ at 20\,cm \citep{nagar99}. 

Our results are shown in Fig.~\ref{fig:N788}. The \feiil\ flux distribution is more extended along the NE-SW direction and presents several knots of enhanced emission. Similar orientations and knots of emission are also observed in Pa$\beta$ and \pii. The ionised gas emission is likely tracing the AGN ionisation structure. We find that the Pa$\beta$ emission is extended mostly along $\rm PA=52^\circ\pm4^\circ$, which is displaced from stellar kinematic PA \citep[$120^\circ\pm3^\circ$;][]{rogemar_stellar} by $\sim68^\circ$. The highest intensity levels in the H$_2$ flux map are more extended along the N--S direction, while at lower flux levels, the H$_2$ emission spreads over the whole NIFS FoV. The \sixl\ is compact and slightly elongated along the AGN ionisation axis, as traced by the orientation of the Pa$\beta$ emission. The \hml/\brg\ vs. \feiil/\pab\ diagnostic diagram of NGC\,788 shows all values in the AGN region, but the \feii/\pab\ ratio is smaller than 2.0 at most locations. The \feii/\pii\  ratio is also smaller than 2.0 at most locations. The Spearman test reveals that there is no correlation between  \hm/\brg\ and \feii/\pab.  The median value of the \feii/\pab\ for the high velocity range is larger than that of the low velocity range.

\subsubsection{Mrk\,607}

Mrk\,607 is an edge-on spiral galaxy ($i=70^{\circ}$) classified as Sa?, hosts a Seyfert 2 nucleus and is located at $\approx 37$\,Mpc \citep{veron-cetty06}.
Hubble Space Telescope (HST) observations show that the [O{\sc iii}] emission is well aligned to the galaxy's major axis ($\rm PA\approx-43^{\circ}$), being more extended along the north-east side 
\citep{ferruit00,schmitt03}. Patches of low stellar velocity dispersion  produced by young to intermediate age stellar populations are observed at distances of 200\,pc from the centre \citep{rogemar_stellar}.
Optical and near-IR IFS observations show that the stellar kinematics %and gas kinematics
in the inner kpc of Mrk\,607 
%are 
is dominated by rotation in the plane of the disc. The gas also presents a clear rotation pattern, but in the opposite direction of the stars %but they are rotating in opposite directions each to other
\citep{Freitas18,rogemar_stellar,astor19}. In addition, equatorial ionised outflows (perpendicular to the ionization cone) are observed in the optical emission lines \citep{Freitas18}. Radio emission is seen only from a nuclear compact source, with no evidence of a radio jet at kpc scales \citep{colbert96,nagar99}.

For Mrk\,607 (Fig.~\ref{fig:M607}), we find that the molecular and ionised gas emission is more extended along the orientation of line of nodes derived from the stellar velocity line of nodes  \citep[$138^\circ\pm3^\circ$;][]{rogemar_stellar}. The ionised gas is more extended to the northwest side of the nucleus and we derive a orientation for the Pa$\beta$ emission  of $143^\circ\pm4^\circ$, consistent with the stellar kinematic PA and with the orientation of optical emission \citep{ferruit00,schmitt03,Freitas18}. The \piil\ and \sixl\ show emission only at distances smaller than 0.8\,arcsec from the nucleus.  As for NGC\,788, the \hm/\brg\ vs. \feii/\pab\ diagnostic diagram shows all values in the AGN region, but a good correlation is found between both ratios.  
The \feii/\pii\  ratio is larger than 2.0 at most locations, reaching values of up to $\sim$7.  Both the  \hm/\brg\ and \feii/\pab\ ratios median values are larger for the high velocity range, compared to the low velocity gas.  Values typical of SF galaxies are seen for the low velocity range in a narrow stripe perpendicular to the galaxy's major axis, co-spatial with the equatorial outflows observed in ionised gas \citep{Freitas18}.

\subsubsection{NGC\,3227}

NGC\,3227 is a spiral galaxy (SAB(s)a), located at a distance of 22\,Mpc and with a Sy 1.5 nucleus \citep{ho97}. A full description of its NIR emission line spectrum is made in \citet{rogerio06}. Radio continuum observations at 18\,cm show two main components separated by 0.4\,arcsec, offset from the optical peak and located at PA$\approx-10^{\circ}$. An outflow is observed in the [O{\sc iii}]$5007$\AA, but it is not aligned with the radio emission \citep{mundell95}. Signatures of outflows are also observed in H$\alpha$ \citep{arribas94} and in H$_2$~1-0S(1) in the central $1-2$\,arcsec \citep{davies14}. 
\citet{schinnerer00} detected an asymmetric nuclear ring of cold molecular gas with 3\,arcsec diameter, which is co-spatial with regions of high \feii/Pa$\beta$ (of up to 4) and H$_2$/Br$\gamma$ (up to 10) values \citep{astor19}. The velocity fields of the hot molecular and ionised gas show a rotation pattern, but it is distorted indicating the presence of additional kinematic components \citep{astor19}.

The flux maps for NGC\,3227  (Fig.~\ref{fig:N3227}), show extended emission over most of the FoV, except for the \sixl, whose emission is restricted to the inner $\sim1$\,arcsec radius. The flux distributions are more elongated along the galaxy's major axis \citep[$156^\circ\pm3^\circ$;][]{rogemar_stellar} and we derive an orientation of $168^\circ\pm6^\circ$ for the Pa$\beta$ emission, which close to the orientation of the radio jet \citep{mundell95}, and thus it is likely tracing the AGN ionisation axis. The \hm/\brg\ vs. \feii/\pab\  diagram shows values in the AGN and HLR regions, with the AGN ratios seen mainly at the centre and close to the borders of the FoV, along the AGN ionisation axis.  
%A good correlation is seen between the two line ratios: 
The \pii/\pab\ line ratio presents values between 2 and 15 in most locations and it correlates with the \feii/\pab. The excitation map for the low velocity range highlights the structure seen in the excitation map using the total fluxes, with values typical of AGN at the nucleus and along the AGN ionisation axis, and HLRs in between. On the other hand, for the high velocity range, most of the ratios fall in the HLR region, except for a region northern of the nucleus, close to the border of the NIFS FoV. As for the previous galaxies, the median line ratios for the high velocity range are  larger than those for the low velocity gas.

\subsubsection{NGC\,3516}

This is a lenticular galaxy, classified as (R)SB0$^0$?(s), with a Sy 1.2 nucleus. Recently a detection of an UV flare placed this galaxy as a changing-look AGN \citep{ilic20}. 
The gas velocity fields deviate from a 
well-behaved rotation pattern, as observed for the near-IR emission lines \citep{astor19} and H$\alpha$ \citep{veilleux93}. This feature is likely due to a bipolar outflow \citep{goad87}  and explained by a precessing twin jet model \citep{veilleux93}.  \citet{astor19} reported  \feii$1.257 \mu$m/Pa$\beta > 2$ in regions with enhanced $\sigma_{\rm[Fe~II]}$ values ($\sim 150$\,km\,s$^{-1}$), indicating shocks. As is the case of Mrk\,607, NGC\,3516 presents patches of smaller stellar velocity dispersion values ($\sim50-80$\,km\,s$^{-1}$) due to the presence of young to intermediate stellar populations. An extended radio emission along $\rm PA\approx8^\circ$ is detected at 20\,cm, while at 3.6\,cm only unresolved nuclear emission is detected \citep{nagar99}. The [O\,{\sc iii}] emission observed by HST shows an S-shaped morphology, extended by 13.6\,arcsec along $\rm PA\approx20^\circ$ \citep{schmitt03}.

The emission-line flux distributions for NGC\,3516 (Fig.~\ref{fig:N3516}) are the most compact among all galaxies in our sample. Only the \feii\ presents emission over the whole FoV. The extended emission is seen mainly along the major axis of the galaxy  \citep[$54^\circ\pm3^\circ$;][]{rogemar_stellar}. We derive an orientation of $51^\circ\pm7^\circ$ for the Pa$\beta$ emission, which may be tracing emission from the galaxy's disc rather than the AGN ionisation structure, as at larger scales, this galaxy presents a clear extended NLR along $\rm PA\approx20^\circ$ \citep{schmitt03}. Not much can be said about the ionisation structure of this galaxy, as we were able to measure all emission lines only at the centre. The observed line ratios fall in the AGN and HLR regions of the diagnostic diagram, and higher median values are found for the line ratios using the high velocity range, as compared to those of the low velocity range.   

\subsubsection{NGC\,5506}

NGC\,5506 is a spiral galaxy classified as Sa\,pec\,edge-on, located at a distance of 31\,Mpc and hosting a Sy 1.9  nucleus  \citep{blanco90,kewley01}. NIFS observations of NGC\,5506 show that \feii$1.257\,\mu$m flux distribution is more extended perpendicularly to the galaxy's major axis ($\Psi_0=90^{\circ}$) \citep{rogemar_stellar,astor19}. This feature, combined with structures in the same spatial location with high velocity dispersion ($\sim 300$\,km\,s$^{-1}$) and the distorted velocity fields, supports the presence of an ionised gas outflow, previously observed using optical long-slit spectra  \citep{maiolino94}. The H$_2\,2.1218 \mu$m is distributed along the major axis and its emission is not as prominent as for the ionised gas, this is probably because the H$_2$ molecule is dissociated by AGN radiation \citep{astor19}.  VLA observations  of NGC\,5506 at 3.6\,cm reveal a linear structure along the east-west direction, surrounded by diffuse emission, with a total extension of 300\,pc \citep{schmitt01}.

The \feii\ flux distribution of NGC\,5506 (Fig.~\ref{fig:N5506}) shows a well defined one-sided cone to the north of the nucleus, which is also seen in the \pii\ flux map. The \hm\ flux distribution is more elongated along the galaxy's major axis \citep[$96^\circ\pm3^\circ$;][]{rogemar_stellar}, presenting a structure that seems to be tracing the outer walls of the cone seen in \feii. The \pab\ shows a round flux distributions and we derive an orientation of   $67^\circ\pm4^\circ$, which seems to trace the east wall of the cone seen in \feii.  The \six\ line emission is slightly more extended to the north of the nucleus.  The  \hm/\brg\ vs. \feii/\pab\ shows values in all regions, with values typical of SF at the nucleus and along the north-south direction, surrounded by the values typical of AGN and HLRs are seen close to the borders of the NIFS FoV. A good correlation is found between  \hm/\brg\ and \feii/\pab, and between  \feii/\pab\ and \feii/\pii, with the latter ratio presenting values of up to 20.  Unlike previous objects, this galaxy presents smaller line ratios for the high velocity gas than for the low velocity range. This behaviour will be further discussed in next section.

\subsubsection{NGC\,5899}

NGC\,5899 is located at 39\,Mpc, classified as SAB(rs)c \citep{ann15} and hosts a Sy\,2 nucleus \citep{devaucouleurs91}.  This galaxy presents a partial ring of low stellar velocity dispersion ($50-80$\,km\,s$^{-1}$) with radius of 200\,pc centred at the galaxy nucleus likely produced by young/intermediate age stellar populations \citep{rogemar_stellar}. 
 \citet{astor19} found that the H$_2\,2.1218 \mu$m emission is extended along the galaxy major axis ($\rm PA=25^{\circ}$) and shows a rotation pattern with small velocity dispersion values over the whole field of view. On the other hand, the strongest \feii\,1.2570\,$ \mu$m and Pa$\beta$ emissions are seen along the north-south direction, as well as the highest gradient in the velocity fields.  The velocity dispersion of the \feii\ reaches values of $\sim 300$\,km\,s$^{-1}$\ reinforcing the hypothesis that the ionised gas traces an outflow, while the molecular gas follows the rotation of the stars \citep{astor19}. To the best of our knowledge, no extended radio emission is detected in NGC\,5899.

The H$_2$ emission in NGC\,5899 (Fig.~\ref{fig:N5899}) is more extended along the orientation of the galaxy's disc \citep[$24^\circ\pm3^\circ$;][]{rogemar_stellar}, while the ionised gas is displaced by 23$^\circ$ from it, and the emission peak of the ionised gas emission is observed at 0.3\,arcsec south of the nucleus. No \sixl\ emission is detected spaxel-by-spaxel, but we were able to measure its flux using integrated spectra, as shown in Tab.\,\ref{tab:ratios}. The excitation map shows values in the AGN region, close to the nucleus, surrounded by HLR regions.  There is no correlation between  \hm/\brg\ and \feii/\pab, but the range of values observed in both line ratios is small. At most locations the  \feii/\pii\ is slightly larger than 2. The median \hm/\brg\ and \feii/\pab\ ratios for the low and high velocity ranges are similar, but the fraction of spaxels in the HLR is larger for the latter.

\subsection{Gas excitation}

The most comprehensive study of the two-dimensional ionisation structure in nearby galaxies using near-IR IFS was performed by \citet{colina15}.  The authors used the \hml/\brg\ and \feii$\,1.6440\,\mu$m/\brg\ diagnostic diagram, as their VLT SINFONI data covers the H and K bands.  The diagnostic diagrams are presented for 10 LIRGs and are compared with results from previously published data of SF and Seyfert galaxies. The results in \citet{colina15} indicates that \hml/\brg\ and \feii$\,1.6440\,\mu$m/\brg\ are correlated in most objects (e.g. IC\,4687), while no correlations are observed in few objects (e.g. NGC\,3256). As the SINFONI covers a larger FoV than NIFS and the ionisation structure in LIRGS is more complex, most of  objects in the sample of \citet{colina15} show line ratios spanning a wide range of values. Some of the Seyfert galaxies in their work clearly show no correlation between the line ratios involved in the diagnostic diagram \citep[e.g., Mrk\,1157;][]{rogemar_mrk1157}. We find strong correlations between \hml/\brg\ and \feiil/\pab\ for Mrk\,607, NGC\,3227 and NGC\,5506, while no correlations are found for the other galaxies, although a trend of increasing \hm/\brg\ with the increasing of \feii/\pab\ is observed for NGC\,3516 (but the number of points is small) and for the high velocity range of NGC\,788. However, considering that the SINFONI data used in \citet{colina15} covers a larger FoV, their observations are seeing limited and their sample has an average redshift larger than ours, a comparison between individual correlations is not straightforward.

\begin{figure}
    \centering
\includegraphics[width=0.45\textwidth]{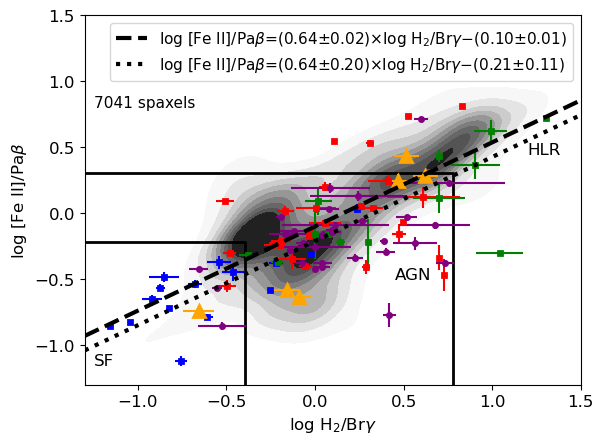} 
\caption{Density plot for the \feiil/\pab\ vs. \hml/\brg\ diagram for our sample. The lines delineating the SF, AGN and HLR regions are from \citet{rogerio13} and the dashed line shows the best linear fit of the data.  The SF, AGN and HLR regions contain 7, 64 and 29 per cent of the points, respectively. The contour levels are equally separated in steps of 10\,\% of the total number of spaxels. The colour points represent line ratios from single aperture spectra compiled from the literature. The squares are from measurements by \citet{larkin98,knop01,reunanen02,dale04,ardila04,ardila05,rogerio06,rogerio13,izotov11} -- blue squares represent SF and blue compact dwarf galaxies, red squares are Seyfert nuclei and green squares represent LINERs. BAT AGN from \citet{lamperti17} are shown as purple circles. Measurements for our sample within 0.5\,arcsec radius are shown as orange triangles. The dotted line represents the best linear fit using the integrated line ratios. } 
    \label{fig:diagnostic}
\end{figure}

\begin{figure}
    \centering
\includegraphics[width=0.45\textwidth]{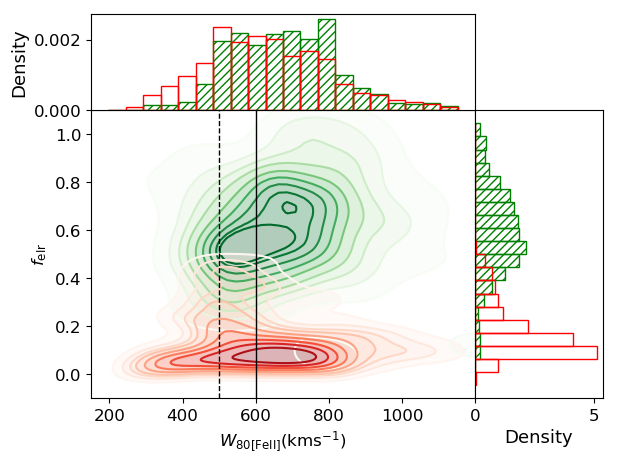}
\caption{Emission-line ratio function vs. $W_{\rm 80 [Fe II]}$ as defined by the Eq.~\ref{elrf}. Red and green contours are from AGN and HLR regions identified in Figs.~\ref{fig:N788} -- \ref{fig:N5899}. The contours are separated in steps of 10\,\% of the total number of spaxels. The vertical lines identify the lower limits of $W_{\rm 80}$ (500 and 600 km\,s$^{-1}$) in ionised outflows.}
    \label{fig:w80}
\end{figure}

We use the emission-line ratios of all galaxies of our sample to construct the \feiil/\pab\ vs. \hml/\brg\  diagnostic diagram (shown in Figure~\ref{fig:diagnostic}), which is useful to investigate the origin of the near-IR emission lines.  We find a strong correlation between \feii/\pab\ and H$_2$/\brg\ in good agreement with previous results using single aperture \citep[e.g.][]{ardila05,rogerio13} and integral field \citep[e.g.][]{colina15} spectra of nearby active galaxies. As the \pab\ and \brg\ emission-lines are originated from the same process (i.e. photoionisation by the AGN or young-massive stars), a possible interpretation of this correlation is that the H$_2$ and \feii\ emissions originate from the same excitation mechanism. 

We use a Spearman correlation test and find a correlation coefficient between  \feii/\pab\  and H$_2$/\brg\ of $R_p=0.68$ and a correlation confidence level larger than 99.9\,\%. 
%We also tested whether these ratios correlate in individual galaxies and find that there is no correlation at a 95\,\% confidence level for only for NGC\,3516, for NGC\,788  and NGC\,5899 there is a weak anti-correlation, while for the other three galaxies the \feii/\pab\  and H$_2$/\brg\ are correlated.As seen in Fig.~\ref{fig:diagnostic}, most of the line ratios are in the AGN region (64\,\% of the spaxels) of the diagram, with a significant fraction in the HLR region (29\,\%) and only a few spaxels (7\,\%) are in the SF region of the diagram. 
This suggests that the \feii\ and H$_2$ emissions in our sample are dominated by thermal processes.  We fit the data by a linear equation resulting in 
 \begin{equation}
 \log \left({\rm \frac{[Fe\,II]}{Pa\beta}}\right) = {\rm(0.65\pm0.026)\times\log \left(\frac{H_2}{Br\gamma}\right) - (0.12\pm0.013)}. 
 \end{equation}
 The best fit equation to our data is shown as a dashed line in Fig.~\ref{fig:diagnostic}, 
 which is consistent with the one obtained by \citet{rogerio13} using single aperture spectra of 67 nearby galaxies [$\log {\rm \frac{[Fe\,II]}{Pa\beta} = (0.749\pm0.072)\times\log \frac{H_2}{Br\gamma} - (0.207\pm0.046)}$].
 
 We have also compiled the \feii/\pab\  and H$_2$/\brg\  ratios measured by integrated spectra  for a sample of 93 objects composed of Star Forming Galaxies \citep{larkin98,dale04,ardila05,rogerio13}, Blue Compact Dwarf Galaxies \citep{izotov11}, Seyfert galaxies \citep{knop01,reunanen02,ardila04,ardila05,rogerio06}, LINERs \citep{larkin98,rogerio13} and BAT AGN \citep{lamperti17}. We show these measurements as colour points in Fig.~\ref{fig:diagnostic} and include the nuclear ratios observed in our sample, within 0.5\,arcsec radius, represented by the orange triangles. As already noticed in \citet{rogerio13}, there is a strong correlation between the nuclear \feii/\pab\  and H$_2$/\brg\ ratios. The Spearman test results in  $R_p=0.66$ and $p-{\rm value}=9.6\times10^{-14}$.  It is worth mentioning that the correlation coefficients for the integrated and spatially resolved line ratios are very similar, meaning that the  \feii/\pab\  and H$_2$/\brg\  diagnostic diagram, originally designed for single aperture measurements, can be used to obtain information on the excitation of the spatially resolved emitting gas. The best linear fit of the single aperture data, shown as a dotted line Fig. \ref{fig:diagnostic}, is well consistent with the resulting fit of the spatially resolved data (dashed line).

Except for NGC\,5506, all galaxies show higher line ratios in the high velocity  range ($100\,{\rm km\,s^{-1}}<|v|<750\,{\rm km\,s^{-1}}$) than in the low velocity range $|v|<100\,{\rm km\,s^{-1}}$). This indicates that shocks contribute to the \feii\ and H$_2$ emission, even in galaxies with most of the spaxels in the AGN region of the diagnostic diagrams (e.g. NGC\,788 and Mrk\,607).

In order to further investigate the origin of the line emission we use two quantities that can be obtained from our data: $W_{80}$ and $f_{\rm elr}$. $W_{80}$ is the width of an emission-line 
above which 80 per cent of the flux is emitted, and is frequently used to identify ionised outflows \citep{zakamska14,wylezalek17,wylezalek20,rogemar_N1275,kakkad20}.
The $W_{80}$ parameter is more sensitive to the wings of the emission-line profiles, and thus a better indicator of outflows than the velocity dispersion ($\sigma$) from a Gaussian function.
Usually  $W_{\rm 80}$ values larger than 600\,km\,s$^{-1}$ in the [O\,{\sc iii}]5007\AA\ line in powerful AGN are associated to outflows \citep[e.g.][]{kakkad20}, while in low-luminosity AGN $W_{\rm 80}\gtrsim{\rm 500\,km\,s^{-1}}$ have been attributed to ionised outflows \citep[e.g.][]{wylezalek20}. We follow a similar procedure used in the optical by \citet{dagostino19b} to separate the contributions of star formation, shocks, and AGN ionisation \citep[see also ][]{dagostino19a} and define an emission-line ratio function ($f_{\rm elr}$) by
 
 \begin{equation}\label{elrf}
   f_{\rm elr} = \frac{\log X-\log X_{\rm P02}}{\log X_{\rm P98}-\log X_{\rm P02}}\times\frac{\log Y-\log Y_{\rm P02}}{\log Y_{\rm P98}-\log Y_{\rm P02}},
 \end{equation}
 with $X=$\hml/\brg\ and $Y=$\feiil/\pab, and P02 and P98 being the percentile 2 and 98 of the observed emission-line ratios, respectively. \citet{dagostino19b} define the emission-line ratio function using [O\,{\sc iii}]5007/H$\beta$ and [N\,{\sc ii}]6583/H$\alpha$ line ratios in the optical and instead of use the P02 and P98 parameters, they use the minimum and maximum ratios. The P02 and P98 parameters are less sensitive to outliers than the minimum and maximum ratio values. 

 We present, in Figure~\ref{fig:w80}, the density plot of $f_{\rm elr}$ vs. $W_{80}$ (corrected for instrumental broadening) for all spaxels with \hml/\brg\ \feiil/\pab\ measurements from all galaxies.   In green, we show the line ratios from the HLR region in the diagnostic diagram  (Figs.~\ref{fig:N788} -- \ref{fig:N5899}) and in red those from the AGN region. We do not show contours for the SF region as they are seen only in one galaxy.
  We find  $W_{\rm 80 [Fe II]}$  larger than the typical limits above which they can be attributed to outflows. The $W_{\rm 80 [Fe II]}$ values derived in this work are larger than those obtained from the $\sigma$ maps presented in \citet{astor19} using the relation: $W80 = 2.563\sigma$ \citep{zakamska14}.  However, \cite{astor19} fitted the emission lines by Gauss-Hermite series the relation above is  no longer valid. In addition, the $h_3$ and $h_4$ maps from \citet{astor19} present high values (with absolute values of up to 0.3), a clear indication that the line profiles are not Gaussian.   A detailed discussion about the gas kinematics will be presented in a forthcoming paper (Bianchin et al., in prep.), but we find ionised outflows in all galaxies of our sample, based on the analysis of multi-Gaussian component fits to the near-IR emission lines.  The interaction of these outflows with the interstellar medium can produce the shocks necessary to excite the H$_2$ and \feii\ emission lines.

In Fig.~\ref{fig:w80}, $f_{\rm elr}$ clearly separates two excitation regimes: the AGN  regime, with $f_{\rm elr}\le0.3$, and the HLR excitation regime for which $f_{\rm elr}$ is higher. In addition, in the HLR excitation regime, $f_{\rm elr}$ increases with $W_{\rm 80 [Fe II]}$, supporting the dominance of shock excitation. A similar increase of  the emission-line ratio function with the gas velocity dispersion  is seen in the optical and associated to shocks \citep{dagostino19a,dagostino19b,rogemar20_letter}. This suggests a strong contribution of shocks to the H$_2$ and \feii\ emission from regions with high \feii/\pab\ and H$_2$/\brg\ ratios in the galaxies of our sample. 
 
 Spaxels in the AGN region of the \feii/\pab\ vs. H$_2$/\brg\ diagram present a wide range in  $W_{\rm 80 [Fe II]}$ and a smaller range of values of $f_{\rm elr}$, as compared to the shock dominated spaxels.  %Although some spaxels in the AGN region present low values of $W_{\rm 80 [Fe II]}$, in most regions the velocities are larger than the limits usually adopted to identify ionised outflows. However, 
 There is no  trend between $f_{\rm elr}$ and $W_{\rm 80 [Fe II]}$ in the red contours of Fig.~\ref{fig:w80}.  This result indicates that shocks are less efficient in the production of line emission, when competing with the AGN radiation field, as found in the optical \citep{zakamska14}. Thus, although a large fraction of the gas is outflowing and outflows naturally produce shocks, the AGN radiation field is the main responsible for the production of the H$_2$ and \feii\ in the AGN region of the NIR diagnostic diagram.   Some contribution of shocks in the AGN region of the diagnostic diagram is supported by the higher line ratios seen in the high velocity range, as discussed above. 

 %  Indeed, we do not find a correlation between the median values of $f_{\rm elr}$ and the hard X-ray luminosity of our galaxies, which could indicate that X-rays do not play an important role in the excitation of the \feii\ and H$_2$ lines. However, such interpretation must be taken with caution as our sample is too small and there are no spatially resolved X-ray images of our galaxies to draw a firm conclusion about the role of X-rays in the production of the \feii\ and H$_2$ emission lines.  

\subsubsection{The origin of the H$_2$ emission}

The near-IR H$_2$ emission lines in the AGN spectra originate from thermal processes \citep{fischer87,moorwood90,veilleux97,ardila05,dors12,rogerio13}. The \feii/\pab\  vs. H$_2$/\brg\ diagnostic diagrams of all galaxies show that most of the line ratios are  in the AGN and HLR regions. This indicates that thermal processes are the most relevant excitation mechanisms in our sample.

The H$_2$ vibrational and rotational temperatures are similar if the gas is in local thermal equilibrium (excited by thermal processes), while for fluorescent excitation, the rotational temperature is expected to be approximately 10 per cent of the vibrational temperature \citep{Sternberg89,ardila04}. We measure both $T_{\rm vib}$ and $T_{\rm rot}$ for NGC\,788, NGC\,3227 and NGC\,5506 (see Fig.~\ref{fig:temperarutes} and Tab.~\ref{tab:temp}). For the these galaxies, the rotational temperatures are smaller than the vibrational, but larger than the values expected for fluorescent excitation. The largest difference is seen in NGC\,788, for which the vibrational temperature is about twice the rotational temperature. For NGC\,5506 the vibrational and rotational temperatures are the most similar, with $T_{\rm vib}\approx 1.4T_{\rm rot}$ within the inner 100 pc of the galaxy and $T_{\rm vib} \approx T_{\rm rot}$ within the 0.5\,arcsec.  This indicates that thermal processes are dominant in all galaxies, but there is also some contribution from non-thermal processes, mainly in NGC\,788. 
%This indicates that non-thermal processes contribute to the observed H$_2$ emission in NGC\,788, while thermal processes are dominant in NGC\,3227 and NGC\,5506.

The H$_2$ 2--1 S(1)/1--0 S(1) vs. 1--0 S(2)/1--0 S(0) diagnostic diagrams (Fig.~\ref{fig:diagrams}) confirm the conclusion that thermal processes dominate the H$_2$ emission in NGC\,788, NGC\,3227, NGC\,5506. Most of the points in the diagrams of Fig.~\ref{fig:diagrams} lie close to the range of values predicted by shock and X-ray excitation models, but some points are also seen close to region predicted for UV excitation of the H$_2$ in dense photon-dominated regions with cloud densities between 10$^4$ and 10$^5$ cm$^{-3}$, exposed to UV radiation field of of young 1--5 Myr star clusters \citep{davies03,davies05}.

We note the points from SF region of the  \feiil/\pab\  vs. \hml/\brg\ diagnostic diagram (Fig.~\ref{fig:N5506}) for NGC\,5506 are seen distant from the region expected for non-thermal excitation and fall very close to the regions predicted by X-ray and shock models in the H$_2$ diagnostic diagram of Fig.~\ref{fig:diagrams}. The H$_2$ and ionised gas emission-line flux distributions are seen almost perpendicular to each other, with the strongest emission in ionised gas being along the north-south direction, at the same locations classified as SF in the excitation map. The [O\,{\sc iii}]5007\AA ~ is also more collimated along the north-south direction and its kinematics is consistent with a bipolar outflow \citep{fischer13}. This suggests the lower \feiil/\pab\  vs. \hml/\brg ~  line ratios along the north-south direction are not produced by star formation, but they may be a consequence of the AGN radiation field. The lower line ratios can be explained if the H$_2$ gas is partially dissociated by the AGN radiation or by shocks from the outflow, decreasing its abundance and consequently its emission, as found for other Seyfert galaxies \citep{sbN4151Exc,rogemar_n1068,Gnilka20}. Similarly, the \feii\ abundance may be reduced by a strong radiation field, favouring higher excitation levels of the iron.

Although we were not able to measure the H$_2$ rotational temperatures and produce the H$_2$ 2--1 S(1)/1--0 S(1) vs. 1--0 S(2)/1--0 S(0) excitation diagrams for Mrk\,607, NGC\,3516 and NGC\,5899, the comparison of their \feiil/\pab\  vs. \hml/\brg\ diagnostic diagram with those of the other galaxies is useful to understand how the H$_2$ emission originates in these objects. Mrk\,607 and NGC\,788 present very similar  \feiil/\pab\  vs. \hml/\brg\ diagrams, with all points in the AGN region, but overall small \feii/\pab. This, together with the fact that the emission of both ionised and molecular gas in Mrk\,607 are similar and observed mainly along the major axis of the galaxy, suggest that we cannot rule out a contribution of fluorescent excitation to the H$_2$ lines. NGC\,3516 and NGC~5899 show high values of both   \feiil/\pab\  and \hml/\brg, suggesting thermal processes are dominant in these galaxies. 

%As higher values of the emission-line ratio function are associated to ionised outflows (Fig.~\ref{fig:w80}), shocks produced by these outflows seem to be the main thermal excitation mechanism, but we cannot rule out the contribution of X-rays, due to the lack of spatially X-ray resolved observations. 
%In summary, in four galaxies of our sample the H$_2$ emission originates mainly in thermal processes, while for two galaxies a contribution of fluorescence to the H$_2$ excitation is supported. 

\subsubsection{The origin of the [Fe\,{\sc ii}] emission}

The \feii\ emission lines in AGN can be produced by photoionisation and shocks. The photoionisation models of \citet{dors12} and \citet{rogerio13} show that an increase in the \feii/\pab\ can originate from an enhancement of the Fe/O abundance. This suggests the correlation seen in Fig.~\ref{fig:diagnostic} could originate in a combination of the amount of ionising photons and a variation in the iron abundance. However, the \feii\ emission can also be increased by shock excitation and indeed, the presence of shocks is supported by the high values of $W_{\rm 80 [Fe II]}$ (Fig.~\ref{fig:w80}). 

The \feii\ and [P\,{\sc ii}] have similar excitation temperatures and ionisation potentials (16.2 and 19.8 eV, respectively) and the \feiil/\piil\ can be used to investigate the origin of the \feii\ emission. The iron is usually locked into dust grains and fast shocks can release it. In photoionised objects, such as the Orion Nebulae, \feiil/\piil$\approx$2 and larger values indicate shocks have destroyed the dust grains, releasing the iron and increasing its abundance \citep{oliva01,jackson07,sbN4151Exc,rogemarMrk1066exc,rogemar_N1275}. In all galaxies, but NGC\,788, most of the spaxels present \feiil/\piil$>$2 indicating shocks play an important role in the production of the \feii\ emission in our sample. In shock dominated objects, values of up to 20 are expected for this line ratio \citep[e.g.][]{oliva01}. Such high values are observed only in some spaxels for NGC\,5506, indicating the \feii\ emission in the galaxies of our sample originates by a combination of photoionisation and shocks. The comparison of the \feii/\pii\ vs. \feii/\pab\ and  \hm/\brg\ vs. \feii/\pab\ (Figs.~\ref{fig:N788} -- \ref{fig:N5899}) shows that high values of  \hm/\brg\ and \feii/\pab\ are not a sufficient condition for shock excitation, although higher values of these ratios usually imply in larger \feii/\pii\ values (e.g. for NGC\,3227 and NGC\,5506).  The shocks necessary to release the iron from the dust grains may be produced by the interaction of the outflows with the ambient gas, as indicated by the high $W_{\rm 80 [Fe II]}$ values.

\subsection{Coronal line emission}

Coronal lines (CLs) are those produced by high ionisation species, with ionisation potentials typically larger than 100\,eV. In the near-IR spectra of AGN, the most common CLs are [S\,{\sc viii}]0.9914\,$\mu$m,  \sixl, [Si\,{\sc x}]1.4305\,$\mu$m, [Si {\sc vi}]1.9650\,$\mu$m, [Al\,{\sc ix}]2.0450\,$\mu$m, [Ca\,{\sc viii}]2.3213\,$\mu$m and [Si \,{\sc vii}]\,2.4830\,$\mu$m. \citep{ardila11}.
 The high ionisation gas in active galaxies can originate from  photoionisation due the AGN ionising continuum \citep[e.g.][]{shields75,korista89,ferguson97}, shocks due to the interaction of a radio jet or outflows with the ambient gas \citep[e.g.][]{osterbrock65,ardila20} or by a combination of photoionisation and shocks \citep[e.g.][]{viegas89,contini01}. 
 Photoionisation seems to be the dominant excitation mechanism of  CLs, but a contribution of shocks is necessary to reproduce the observed line intensities, specially at distances larger than $\sim$100\,pc from the nucleus \citep{ardila06,ardila11,ardila20,geballe09}.

The CLs usually trace highly ionised AGN-driven outflows, as indicated by their broader and blueshifted profiles as compared to those from lower ionisation gas \citep{ardila02,ardila06,ms11,ardila17}.  For example, in the Circinus galaxy, highly ionised gas (traced by the [Fe\,{\sc vii}]6087\,$\mu$m) is seen up to 700\,pc from the nucleus, and is tracing shock ionisation due to the interaction of the radio jet with the ambient gas \citep{ardila20}. 

Our NIFS data provides a unique capability to map the extension of the coronal line emission in nearby galaxies. The most prominent CL in 
in our observed spectral window is \sixl. The \sixl\ flux distribution is spatially resolved in all galaxies of our sample, except for NGC\,5899, which only presents a very faint nuclear [S\,{\sc ix}] emission. Table~\ref{tab:CLR} lists the full width at half maximum (FWHM$_{\rm S\,{IX}}$) of the [S\,{\sc ix}] flux distributions and the most extended [S\,{\sc ix}] emission at 3$\sigma$ level detection ($R_{\rm S\,{IX}}$) for the galaxies of our sample. We find FWHM$_{\rm S\,{IX}}$ ranging from 30 to 90 pc, and  $R_{\rm S\,{IX}}$ in the range 80 -- 185 pc. The extension of the [S\,{\sc ix}] in our sample is slightly larger than those found by \citet{ms11} using IFS of a sample of seven nearby Seyfert galaxies. They used the [Si {\sc vi}]1.9650\,$\mu$m emission as a tracer of the coronal line region and found FWHM from 8 to 60\,pc and sizes from 8 to 150\,pc. \citet{prieto05} derived extensions of the [Si\,{\sc vii}]2.48\,$\mu$m from 30 to 200\,pc using adaptive optics narrow band images with the ESO/VLT. At these scales, a pure photoionisation scenario fails to reproduce the CLs intensities, favoring an additional contribution of shocks to the coronal emission \citep[e.g.][]{ardila06,mazzalay13,ardila20}.  
Most of galaxies of our sample show compact radio emission \citep{nagar99,schmitt01}. The only exception is NGC\,3227, which shows extended radio emission along $\rm PA\approx-10^\circ$ \citep{mundell95}. However, the CL emission in this galaxy does not present a clear extension along the direction of the radio structure. On the other hand,  ionised outflows are common in our sample (Bianchin et al., in prep.) and thus, 
 the most likely scenario to produce CLs observed in our sample is the combination of photoionisation and shocks due to AGN-driven outflows \citep[e.g.][]{viegas89,contini01}.

%Si VI 205.279
%S IX  379.84
%Si VI  246.57
%MS Circinus, NGC 1068, NGC 2992, NGC 3783, NGC 4151, NGC 6814, NGC 7469
%Prieto Circinus, NGC 1068, ESO 428-G014, NGC 3081

\begin{table}
	\centering
	\caption{Extensions of the \sixl\ emission in our sample. FWHM$_{\rm S\,IX}$ is the full width at half maximum of the \sixl\ flux distribution in arcseconds and parsecs, $R_{\rm S\,IX}$ is distance to the nucleus of the most extended [S\,{\sc ix}] emission and FWHM$_{\rm PSF}$ is the J band PSF \citep{rogemar_sample}.  }
	\label{tab:CLR}
	\begin{tabular}{lcccc} % four columns, alignment for each
\hline
Galaxy & FWHM$_{\rm S\,IX}$ &  FWHM$_{\rm S\,IX}$ & $R_{\rm S\,IX}$ & FWHM$_{\rm PSF}$  \\
 & (arcsec) & (pc) & (pc) & (arcsec)\\
\hline
NGC788 & 0.32 & 90 &  185 & 0.13 \\
Mrk607 & 0.34 & 63 &  150 & 0.14 \\ 
NGC3227 & 0.36 & 30 & 80 & 0.13 \\
NGC3516 & 0.37 & 68 & 120 & 0.17 \\ 
NGC5506 & 0.47 & 60 & 100 & 0.15 \\ 	
	\hline
	\end{tabular}
\end{table}

\section{Conclusions}
\label{sec:conclusions}

We use Gemini NIFS J and K band data to investigate the origin of the near-IR emission lines in the inner few hundred pc of six luminous Seyfert galaxies (NGC\,788, Mrk\,607, NGC\,3227, NGC\,3516, NGC\,5506 and NGC\,5899) at spatial resolutions of 10 to 40 pc. Our main conclusions are the following:

\begin{itemize}
    \item For all galaxies, the \feiil/\pab\  vs. \hml/\brg\ diagrams present values in the AGN and HLR regions at most locations. We find a strong correlation between \feiil/\pab\  and \hml/\brg\ represented by the linear equation $\log ({\rm [Fe\,II]/Pa\beta) = (0.65\pm0.026)\times\log({H_2/Br\gamma)} - (0.12\pm0.013)}$ which is consistent with the previous result of \citet{rogerio13},  possibly indicating a common excitation origin.
    
    \item  The HLR regions in the \feiil/\pab\  vs. \hml/\brg\ diagnostic diagram are observed surrounding the AGN region, with no preferential orientation relative to the orientation of the ionisation structure.
    
    \item We define an emission-line ratio function ($f_{\rm elr}$) that depends on the  \feiil/\pab\  vs. \hml/\brg\ line ratios and presents values in the range $\sim$0--1. We find a correlation between this function and $W_{\rm 80 [Fe II]}$, where $f_{\rm elr}$ increases with $W_{\rm 80 [Fe II]}$ for the HLR regions of the galaxies indicating shocks produced by gas outflows contribute to the line emission  in locations with high \feiil/\pab\  and \hml/\brg\ values.

    \item The H$_2$ emission lines are excited by thermal processes in all galaxies. Based on line-ratio diagnostic diagrams and the trend seen between $f_{\rm elr}$ and the $W_{\rm 80 [Fe II]}$, our results indicate, for spaxels in the HLR region, the main process exciting the H$_2$ lines are shocks produced by AGN-driven winds. While for spaxels in the AGN region of the diagnostic diagram, the AGN radiation field is responsible for the observed H$_2$ emission. A small contribution of shocks cannot be ruled out in the AGN region, as indicated by the higher line ratios obtained by computing the line fluxes at velocities of $|v|>100$\,km\,s$^{-1}$, compared to those using $|v|<100$\,km\,s$^{-1}$, relative to the peak velocity.
    
  %  \item The mean excitation temperatures of the H$_2$ emitting gas are in the range $\sim2400$ to $\sim5200$ K, with the highest value observed for NGC\,788.
    
   \item The \feii\ emission in our sample originates from a combination of photoionisation by the central source and shocks due to AGN winds, as indicated by the \feiil/\piil\ high values (2--20), line ratios and the correlation between $f_{\rm elr}$ and the $W_{\rm 80 [Fe II]}$ for the HLR region. 
   
    \item The coronal line region, traced by \sixl\ emission, extends out between 80 and 185 pc from the galaxy nucleus in five galaxies and is likely produced by AGN photoionisation. The \sixl\ emission line is only detected in NGC\,5899 using an integrated spectrum. 
\end{itemize}

\section*{Acknowledgements}
We thank an anonymous referee for useful suggestions which helped to improve the paper. 
This study was financed in part by Conselho Nacional de Desenvolvimento Cient\'ifico e Tecnol\'ogico (202582/2018-3, 304927/2017-1 and 400352/2016-8) and Funda\c c\~ao de Amparo \`a pesquisa do Estado do Rio Grande do Sul (17/2551-0001144-9 and 16/2551-0000251-7). M.B. thanks the financial support from Coordenação de Aperfeiçoamento de Pessoal de Nível Superior - Brasil (CAPES) - Finance Code 001.  R.R. thanks CNPq, CAPES and FAPERGS for financial support. N.Z.D. acknowledges partial support from FONDECYT through project 3190769.
Based on observations obtained at the Gemini Observatory, which is operated by the Association of Universities for Research in Astronomy, Inc., under a cooperative agreement with the NSF on behalf of the Gemini partnership: the National Science Foundation (United States), National Research Council (Canada), CONICYT (Chile), Ministerio de Ciencia, Tecnolog\'{i}a e Innovaci\'{o}n Productiva (Argentina), Minist\'{e}rio da Ci\^{e}ncia, Tecnologia e Inova\c{c}\~{a}o (Brazil), and Korea Astronomy and Space Science Institute (Republic of Korea). 
This research has made use of NASA's Astrophysics Data System Bibliographic Services. This research has made use of the NASA/IPAC Extragalactic Database (NED), which is operated by the Jet Propulsion Laboratory, California Institute of Technology, under contract with the National Aeronautics and Space Administration. The Pan-STARRS1 Surveys (PS1) and the PS1 public science archive have been made possible through contributions by the Institute for Astronomy, the University of Hawaii, the Pan-STARRS Project Office, the Max-Planck Society and its participating institutes, the Max Planck Institute for Astronomy, Heidelberg and the Max Planck Institute for Extraterrestrial Physics, Garching, The Johns Hopkins University, Durham University, the University of Edinburgh, the Queen's University Belfast, the Harvard-Smithsonian centre for Astrophysics, the Las Cumbres Observatory Global Telescope Network Incorporated, the National Central University of Taiwan, the Space Telescope Science Institute, the National Aeronautics and Space Administration under Grant No. NNX08AR22G issued through the Planetary Science Division of the NASA Science Mission Directorate, the National Science Foundation Grant No. AST-1238877, the University of Maryland, Eotvos Lorand University (ELTE), the Los Alamos National Laboratory, and the Gordon and Betty Moore Foundation. 

%%%%%%%%%%%%%%%%%%%%%%%%%%%%%%%%%%%%%%%%%%%%%%%%%%

\section*{Data availability}
The data used in this work are publicly available online via the GEMINI archive https://archive.gemini.edu/searchform, under the following program codes: GN-2012B-Q-45, GN-2013A-Q-48, GN-2015A-Q-3, GN-2015B-Q-29 and GN-2016A-Q-6.  The processed datacubes used in this article will be shared on reasonable request to the corresponding author.

%%%%%%%%%%%%%%%%%%%% REFERENCES %%%%%%%%%%%%%%%%%%%%%%%%%%%%%%%%%%%%%%%%%%%%

\bibliographystyle{mnras}
\bibliography{paper}

\begin{thebibliography}{}
\makeatletter
\relax
\def\mn@urlcharsother{\let\do\@makeother \do\$\do\&\do\#\do\^\do\_\do\%\do\~}
\def\mn@doi{\begingroup\mn@urlcharsother \@ifnextchar [ {\mn@doi@}
  {\mn@doi@[]}}
\def\mn@doi@[#1]#2{\def\@tempa{#1}\ifx\@tempa\@empty \href
  {http://dx.doi.org/#2} {doi:#2}\else \href {http://dx.doi.org/#2} {#1}\fi
  \endgroup}
\def\mn@eprint#1#2{\mn@eprint@#1:#2::\@nil}
\def\mn@eprint@arXiv#1{\href {http://arxiv.org/abs/#1} {{\tt arXiv:#1}}}
\def\mn@eprint@dblp#1{\href {http://dblp.uni-trier.de/rec/bibtex/#1.xml}
  {dblp:#1}}
\def\mn@eprint@#1:#2:#3:#4\@nil{\def\@tempa {#1}\def\@tempb {#2}\def\@tempc
  {#3}\ifx \@tempc \@empty \let \@tempc \@tempb \let \@tempb \@tempa \fi \ifx
  \@tempb \@empty \def\@tempb {arXiv}\fi \@ifundefined
  {mn@eprint@\@tempb}{\@tempb:\@tempc}{\expandafter \expandafter \csname
  mn@eprint@\@tempb\endcsname \expandafter{\@tempc}}}

\bibitem[\protect\citeauthoryear{{Ann}, {Seo}  \& {Ha}}{{Ann}
  et~al.}{2015}]{ann15}
{Ann} H.~B.,  {Seo} M.,   {Ha} D.~K.,  2015, \mn@doi [The Astrophysical Journal
  Supplement Series] {10.1088/0067-0049/217/2/27}, \href
  {https://ui.adsabs.harvard.edu/abs/2015ApJS..217...27A} {217, 27}

\bibitem[\protect\citeauthoryear{{Arribas} \& {Mediavilla}}{{Arribas} \&
  {Mediavilla}}{1994}]{arribas94}
{Arribas} S.,  {Mediavilla} E.,  1994, \mn@doi [Astrophysical Journal]
  {10.1086/174983}, \href
  {https://ui.adsabs.harvard.edu/abs/1994ApJ...437..149A} {437, 149}

\bibitem[\protect\citeauthoryear{{Black} \& {van Dishoeck}}{{Black} \& {van
  Dishoeck}}{1987}]{black87}
{Black} J.~H.,  {van Dishoeck} E.~F.,  1987, \mn@doi [\apj] {10.1086/165740},
  \href {https://ui.adsabs.harvard.edu/abs/1987ApJ...322..412B} {322, 412}

\bibitem[\protect\citeauthoryear{{Blanco}, {Ward}  \& {Wright}}{{Blanco}
  et~al.}{1990}]{blanco90}
{Blanco} P.~R.,  {Ward} M.~J.,   {Wright} G.~S.,  1990, \mn@doi [Monthly
  Notices of the Royal Astronomical Society] {10.1093/mnras/242.1.4P}, \href
  {https://ui.adsabs.harvard.edu/abs/1990MNRAS.242P...4B} {242, 4P}

\bibitem[\protect\citeauthoryear{{Colbert}, {Baum}, {Gallimore}, {O'Dea}  \&
  {Christensen}}{{Colbert} et~al.}{1996}]{colbert96}
{Colbert} E. J.~M.,  {Baum} S.~A.,  {Gallimore} J.~F.,  {O'Dea} C.~P.,
  {Christensen} J.~A.,  1996, \mn@doi [Astrophysical Journal] {10.1086/177633},
  \href {https://ui.adsabs.harvard.edu/abs/1996ApJ...467..551C} {467, 551}

\bibitem[\protect\citeauthoryear{{Colina} et~al.,}{{Colina}
  et~al.}{2015}]{colina15}
{Colina} L.,  et~al., 2015, \mn@doi [\aap] {10.1051/0004-6361/201425567}, \href
  {https://ui.adsabs.harvard.edu/abs/2015A&A...578A..48C} {578, A48}

\bibitem[\protect\citeauthoryear{{Contini} \& {Viegas}}{{Contini} \&
  {Viegas}}{2001}]{contini01}
{Contini} M.,  {Viegas} S.~M.,  2001, \mn@doi [\apjs] {10.1086/318956}, \href
  {https://ui.adsabs.harvard.edu/abs/2001ApJS..132..211C} {132, 211}

\bibitem[\protect\citeauthoryear{{D'Agostino}, {Kewley}, {Groves}, {Medling},
  {Dopita}  \& {Thomas}}{{D'Agostino} et~al.}{2019a}]{dagostino19a}
{D'Agostino} J.~J.,  {Kewley} L.~J.,  {Groves} B.~A.,  {Medling} A.,  {Dopita}
  M.~A.,   {Thomas} A.~D.,  2019a, \mn@doi [\mnras] {10.1093/mnrasl/slz028},
  \href {https://ui.adsabs.harvard.edu/abs/2019MNRAS.485L..38D} {485, L38}

\bibitem[\protect\citeauthoryear{{D'Agostino} et~al.,}{{D'Agostino}
  et~al.}{2019b}]{dagostino19b}
{D'Agostino} J.~J.,  et~al., 2019b, \mn@doi [\mnras] {10.1093/mnras/stz1611},
  \href {https://ui.adsabs.harvard.edu/abs/2019MNRAS.487.4153D} {487, 4153}

\bibitem[\protect\citeauthoryear{{Dale} et~al.,}{{Dale} et~al.}{2004}]{dale04}
{Dale} D.~A.,  et~al., 2004, \mn@doi [\apj] {10.1086/380753}, \href
  {https://ui.adsabs.harvard.edu/abs/2004ApJ...601..813D} {601, 813}

\bibitem[\protect\citeauthoryear{{Davies}, {Sternberg}, {Lehnert}  \&
  {Tacconi-Garman}}{{Davies} et~al.}{2003}]{davies03}
{Davies} R.~I.,  {Sternberg} A.,  {Lehnert} M.,   {Tacconi-Garman} L.~E.,
  2003, \mn@doi [\apj] {10.1086/378634}, \href
  {https://ui.adsabs.harvard.edu/abs/2003ApJ...597..907D} {597, 907}

\bibitem[\protect\citeauthoryear{{Davies}, {Sternberg}, {Lehnert}  \&
  {Tacconi-Garman}}{{Davies} et~al.}{2005}]{davies05}
{Davies} R.~I.,  {Sternberg} A.,  {Lehnert} M.~D.,   {Tacconi-Garman} L.~E.,
  2005, \mn@doi [\apj] {10.1086/444495}, \href
  {https://ui.adsabs.harvard.edu/abs/2005ApJ...633..105D} {633, 105}

\bibitem[\protect\citeauthoryear{{Davies} et~al.,}{{Davies}
  et~al.}{2014}]{davies14}
{Davies} R.~I.,  et~al., 2014, \mn@doi [\apj] {10.1088/0004-637X/792/2/101},
  \href {https://ui.adsabs.harvard.edu/abs/2014ApJ...792..101D} {792, 101}

\bibitem[\protect\citeauthoryear{{Dors}, {Riffel}, {Cardaci}, {H{\"a}gele},
  {Krabbe}, {P{\'e}rez-Montero}  \& {Rodrigues}}{{Dors} et~al.}{2012}]{dors12}
{Dors} Oli~L. J.,  {Riffel} R.~A.,  {Cardaci} M.~V.,  {H{\"a}gele} G.~F.,
  {Krabbe} {\'A}.~C.,  {P{\'e}rez-Montero} E.,   {Rodrigues} I.,  2012, \mn@doi
  [\mnras] {10.1111/j.1365-2966.2012.20600.x}, \href
  {https://ui.adsabs.harvard.edu/abs/2012MNRAS.422..252D} {422, 252}

\bibitem[\protect\citeauthoryear{{Draine} \& {Woods}}{{Draine} \&
  {Woods}}{1990}]{draine90}
{Draine} B.~T.,  {Woods} D.~T.,  1990, \mn@doi [\apj] {10.1086/169358}, \href
  {https://ui.adsabs.harvard.edu/abs/1990ApJ...363..464D} {363, 464}

\bibitem[\protect\citeauthoryear{{Durr{\'e}} \& {Mould}}{{Durr{\'e}} \&
  {Mould}}{2018}]{durre18}
{Durr{\'e}} M.,  {Mould} J.,  2018, \mn@doi [\apj] {10.3847/1538-4357/aae68e},
  \href {https://ui.adsabs.harvard.edu/abs/2018ApJ...867..149D} {867, 149}

\bibitem[\protect\citeauthoryear{{Ferguson}, {Korista}  \&
  {Ferland}}{{Ferguson} et~al.}{1997}]{ferguson97}
{Ferguson} J.~W.,  {Korista} K.~T.,   {Ferland} G.~J.,  1997, \mn@doi [\apjs]
  {10.1086/312998}, \href
  {https://ui.adsabs.harvard.edu/abs/1997ApJS..110..287F} {110, 287}

\bibitem[\protect\citeauthoryear{{Ferruit}, {Wilson}  \& {Mulchaey}}{{Ferruit}
  et~al.}{2000}]{ferruit00}
{Ferruit} P.,  {Wilson} A.~S.,   {Mulchaey} J.,  2000, \mn@doi [The
  Astrophysical Journal Supplement Series] {10.1086/313379}, \href
  {https://ui.adsabs.harvard.edu/abs/2000ApJS..128..139F} {128, 139}

\bibitem[\protect\citeauthoryear{{Fischer}, {Geballe}, {Smith}, {Simon}  \&
  {Storey}}{{Fischer} et~al.}{1987}]{fischer87}
{Fischer} J.,  {Geballe} T.~R.,  {Smith} H.~A.,  {Simon} M.,   {Storey}
  J.~W.~V.,  1987, \mn@doi [\apj] {10.1086/165584}, \href
  {https://ui.adsabs.harvard.edu/abs/1987ApJ...320..667F} {320, 667}

\bibitem[\protect\citeauthoryear{{Fischer}, {Crenshaw}, {Kraemer}  \&
  {Schmitt}}{{Fischer} et~al.}{2013}]{fischer13}
{Fischer} T.~C.,  {Crenshaw} D.~M.,  {Kraemer} S.~B.,   {Schmitt} H.~R.,  2013,
  \mn@doi [The Astrophysical Journal Supplement Series]
  {10.1088/0067-0049/209/1/1}, \href
  {https://ui.adsabs.harvard.edu/abs/2013ApJS..209....1F} {209, 1}

\bibitem[\protect\citeauthoryear{{Forbes} \& {Ward}}{{Forbes} \&
  {Ward}}{1993}]{forbes93}
{Forbes} D.~A.,  {Ward} M.~J.,  1993, \mn@doi [\apj] {10.1086/173221}, \href
  {https://ui.adsabs.harvard.edu/abs/1993ApJ...416..150F} {416, 150}

\bibitem[\protect\citeauthoryear{{Freitas} et~al.,}{{Freitas}
  et~al.}{2018}]{Freitas18}
{Freitas} I.~C.,  et~al., 2018, \mn@doi [\mnras] {10.1093/mnras/sty303}, \href
  {https://ui.adsabs.harvard.edu/abs/2018MNRAS.476.2760F} {476, 2760}

\bibitem[\protect\citeauthoryear{{Geballe}, {Mason}, {Rodr{\'\i}guez-Ardila}
  \& {Axon}}{{Geballe} et~al.}{2009}]{geballe09}
{Geballe} T.~R.,  {Mason} R.~E.,  {Rodr{\'\i}guez-Ardila} A.,   {Axon} D.~J.,
  2009, \mn@doi [\apj] {10.1088/0004-637X/701/2/1710}, \href
  {https://ui.adsabs.harvard.edu/abs/2009ApJ...701.1710G} {701, 1710}

\bibitem[\protect\citeauthoryear{{Glidden}, {Rose}, {Elvis}  \&
  {McDowell}}{{Glidden} et~al.}{2016}]{glidden16}
{Glidden} A.,  {Rose} M.,  {Elvis} M.,   {McDowell} J.,  2016, \mn@doi [\apj]
  {10.3847/0004-637X/824/1/34}, \href
  {https://ui.adsabs.harvard.edu/abs/2016ApJ...824...34G} {824, 34}

\bibitem[\protect\citeauthoryear{{Gnilka} et~al.,}{{Gnilka}
  et~al.}{2020}]{Gnilka20}
{Gnilka} C.~L.,  et~al., 2020, \mn@doi [\apj] {10.3847/1538-4357/ab8000}, \href
  {https://ui.adsabs.harvard.edu/abs/2020ApJ...893...80G} {893, 80}

\bibitem[\protect\citeauthoryear{{Goad} \& {Gallagher}}{{Goad} \&
  {Gallagher}}{1987}]{goad87}
{Goad} J.~W.,  {Gallagher} J. S.~I.,  1987, \mn@doi [Astronomical Journal]
  {10.1086/114499}, \href
  {https://ui.adsabs.harvard.edu/abs/1987AJ.....94..640G} {94, 640}

\bibitem[\protect\citeauthoryear{{Guillard}, {Boulanger}, {Pineau Des
  For{\^e}ts}  \& {Appleton}}{{Guillard} et~al.}{2009}]{guillard09}
{Guillard} P.,  {Boulanger} F.,  {Pineau Des For{\^e}ts} G.,   {Appleton}
  P.~N.,  2009, \mn@doi [\aap] {10.1051/0004-6361/200811263}, \href
  {https://ui.adsabs.harvard.edu/abs/2009A&A...502..515G} {502, 515}

\bibitem[\protect\citeauthoryear{{Ho}, {Filippenko}  \& {Sargent}}{{Ho}
  et~al.}{1997}]{ho97}
{Ho} L.~C.,  {Filippenko} A.~V.,   {Sargent} W. L.~W.,  1997, \mn@doi [The
  Astrophysical Journal Supplement Series] {10.1086/313041}, \href
  {https://ui.adsabs.harvard.edu/abs/1997ApJS..112..315H} {112, 315}

\bibitem[\protect\citeauthoryear{{Ho} et~al.,}{{Ho} et~al.}{2014}]{ho14}
{Ho} I.~T.,  et~al., 2014, \mn@doi [\mnras] {10.1093/mnras/stu1653}, \href
  {https://ui.adsabs.harvard.edu/abs/2014MNRAS.444.3894H} {444, 3894}

\bibitem[\protect\citeauthoryear{{Hollenbach} \& {McKee}}{{Hollenbach} \&
  {McKee}}{1989}]{hollenbach89}
{Hollenbach} D.,  {McKee} C.~F.,  1989, \mn@doi [\apj] {10.1086/167595}, \href
  {https://ui.adsabs.harvard.edu/abs/1989ApJ...342..306H} {342, 306}

\bibitem[\protect\citeauthoryear{{Ichikawa}, {Ricci}, {Ueda}, {Matsuoka},
  {Toba}, {Kawamuro}, {Trakhtenbrot}  \& {Koss}}{{Ichikawa}
  et~al.}{2017}]{ichikawa17}
{Ichikawa} K.,  {Ricci} C.,  {Ueda} Y.,  {Matsuoka} K.,  {Toba} Y.,  {Kawamuro}
  T.,  {Trakhtenbrot} B.,   {Koss} M.~J.,  2017, \mn@doi [\apj]
  {10.3847/1538-4357/835/1/74}, \href
  {https://ui.adsabs.harvard.edu/abs/2017ApJ...835...74I} {835, 74}

\bibitem[\protect\citeauthoryear{{Ili{\'c}} et~al.,}{{Ili{\'c}}
  et~al.}{2020}]{ilic20}
{Ili{\'c}} D.,  et~al., 2020, \mn@doi [Astronomy \& Astrophysics]
  {10.1051/0004-6361/202037532}, \href
  {https://ui.adsabs.harvard.edu/abs/2020A&A...638A..13I} {638, A13}

\bibitem[\protect\citeauthoryear{{Izotov} \& {Thuan}}{{Izotov} \&
  {Thuan}}{2011}]{izotov11}
{Izotov} Y.~I.,  {Thuan} T.~X.,  2011, \mn@doi [\apj]
  {10.1088/0004-637X/734/2/82}, \href
  {https://ui.adsabs.harvard.edu/abs/2011ApJ...734...82I} {734, 82}

\bibitem[\protect\citeauthoryear{{Jackson} \& {Beswick}}{{Jackson} \&
  {Beswick}}{2007}]{jackson07}
{Jackson} N.,  {Beswick} R.~J.,  2007, \mn@doi [\mnras]
  {10.1111/j.1365-2966.2007.11467.x}, \href
  {https://ui.adsabs.harvard.edu/abs/2007MNRAS.376..719J} {376, 719}

\bibitem[\protect\citeauthoryear{{Kakkad} et~al.,}{{Kakkad}
  et~al.}{2020}]{kakkad20}
{Kakkad} D.,  et~al., 2020, arXiv e-prints, \href
  {https://ui.adsabs.harvard.edu/abs/2020arXiv200801728K} {p. arXiv:2008.01728}

\bibitem[\protect\citeauthoryear{{Kewley}, {Heisler}, {Dopita}  \&
  {Lumsden}}{{Kewley} et~al.}{2001}]{kewley01}
{Kewley} L.~J.,  {Heisler} C.~A.,  {Dopita} M.~A.,   {Lumsden} S.,  2001,
  \mn@doi [The Astrophysical Journal Supplement Series] {10.1086/318944}, \href
  {https://ui.adsabs.harvard.edu/abs/2001ApJS..132...37K} {132, 37}

\bibitem[\protect\citeauthoryear{{Knop}, {Armus}, {Matthews}, {Murphy}  \&
  {Soifer}}{{Knop} et~al.}{2001}]{knop01}
{Knop} R.~A.,  {Armus} L.,  {Matthews} K.,  {Murphy} T.~W.,   {Soifer} B.~T.,
  2001, \mn@doi [\aj] {10.1086/322068}, \href
  {https://ui.adsabs.harvard.edu/abs/2001AJ....122..764K} {122, 764}

\bibitem[\protect\citeauthoryear{{Korista} \& {Ferland}}{{Korista} \&
  {Ferland}}{1989}]{korista89}
{Korista} K.~T.,  {Ferland} G.~J.,  1989, \mn@doi [\apj] {10.1086/167739},
  \href {https://ui.adsabs.harvard.edu/abs/1989ApJ...343..678K} {343, 678}

\bibitem[\protect\citeauthoryear{{Kwan}, {Gatley}, {Merrill}, {Probst}  \&
  {Weintraub}}{{Kwan} et~al.}{1977}]{Kwan77}
{Kwan} J.~H.,  {Gatley} I.,  {Merrill} K.~M.,  {Probst} R.,   {Weintraub}
  D.~A.,  1977, \mn@doi [\apj] {10.1086/155514}, \href
  {https://ui.adsabs.harvard.edu/abs/1977ApJ...216..713K} {216, 713}

\bibitem[\protect\citeauthoryear{{Lamperti} et~al.,}{{Lamperti}
  et~al.}{2017}]{lamperti17}
{Lamperti} I.,  et~al., 2017, \mn@doi [\mnras] {10.1093/mnras/stx055}, \href
  {https://ui.adsabs.harvard.edu/abs/2017MNRAS.467..540L} {467, 540}

\bibitem[\protect\citeauthoryear{{Larkin}, {Armus}, {Knop}, {Soifer}  \&
  {Matthews}}{{Larkin} et~al.}{1998}]{larkin98}
{Larkin} J.~E.,  {Armus} L.,  {Knop} R.~A.,  {Soifer} B.~T.,   {Matthews} K.,
  1998, \mn@doi [\apjs] {10.1086/313063}, \href
  {https://ui.adsabs.harvard.edu/abs/1998ApJS..114...59L} {114, 59}

\bibitem[\protect\citeauthoryear{{Lepp} \& {McCray}}{{Lepp} \&
  {McCray}}{1983}]{Lepp83}
{Lepp} S.,  {McCray} R.,  1983, \mn@doi [\apj] {10.1086/161062}, \href
  {https://ui.adsabs.harvard.edu/abs/1983ApJ...269..560L} {269, 560}

\bibitem[\protect\citeauthoryear{{Liu}, {Zakamska}, {Greene}, {Nesvadba}  \&
  {Liu}}{{Liu} et~al.}{2013}]{liu13}
{Liu} G.,  {Zakamska} N.~L.,  {Greene} J.~E.,  {Nesvadba} N. P.~H.,   {Liu} X.,
   2013, \mn@doi [\mnras] {10.1093/mnras/stt1755}, \href
  {https://ui.adsabs.harvard.edu/abs/2013MNRAS.436.2576L} {436, 2576}

\bibitem[\protect\citeauthoryear{{Maiolino}, {Stanga}, {Salvati}  \& {Rodriguez
  Espinosa}}{{Maiolino} et~al.}{1994}]{maiolino94}
{Maiolino} R.,  {Stanga} R.,  {Salvati} M.,   {Rodriguez Espinosa} J.~M.,
  1994, Astronomy and Astrophysics, \href
  {https://ui.adsabs.harvard.edu/abs/1994A&A...290...40M} {290, 40}

\bibitem[\protect\citeauthoryear{{Maloney}, {Hollenbach}  \&
  {Tielens}}{{Maloney} et~al.}{1996}]{maloney96}
{Maloney} P.~R.,  {Hollenbach} D.~J.,   {Tielens} A.~G.~G.~M.,  1996, \mn@doi
  [\apj] {10.1086/177532}, \href
  {https://ui.adsabs.harvard.edu/abs/1996ApJ...466..561M} {466, 561}

\bibitem[\protect\citeauthoryear{{May} \& {Steiner}}{{May} \&
  {Steiner}}{2017}]{may17}
{May} D.,  {Steiner} J.~E.,  2017, \mn@doi [\mnras] {10.1093/mnras/stx886},
  \href {https://ui.adsabs.harvard.edu/abs/2017MNRAS.469..994M} {469, 994}

\bibitem[\protect\citeauthoryear{{May}, {Rodr{\'\i}guez-Ardila}, {Prieto},
  {Fern{\'a}ndez-Ontiveros}, {Diaz}  \& {Mazzalay}}{{May} et~al.}{2018}]{may18}
{May} D.,  {Rodr{\'\i}guez-Ardila} A.,  {Prieto} M.~A.,
  {Fern{\'a}ndez-Ontiveros} J.~A.,  {Diaz} Y.,   {Mazzalay} X.,  2018, \mn@doi
  [\mnras] {10.1093/mnrasl/sly155}, \href
  {https://ui.adsabs.harvard.edu/abs/2018MNRAS.481L.105M} {481, L105}

\bibitem[\protect\citeauthoryear{{May}, {Steiner}, {Menezes}, {Williams}  \&
  {Wang}}{{May} et~al.}{2020}]{may20}
{May} D.,  {Steiner} J.~E.,  {Menezes} R.~B.,  {Williams} D.~R.~A.,   {Wang}
  J.,  2020, \mn@doi [\mnras] {10.1093/mnras/staa1545}, \href
  {https://ui.adsabs.harvard.edu/abs/2020MNRAS.496.1488M} {496, 1488}

\bibitem[\protect\citeauthoryear{{Mazzalay}, {Rodr{\'\i}guez-Ardila}, {Komossa}
   \& {McGregor}}{{Mazzalay} et~al.}{2013}]{mazzalay13}
{Mazzalay} X.,  {Rodr{\'\i}guez-Ardila} A.,  {Komossa} S.,   {McGregor} P.~J.,
  2013, \mn@doi [\mnras] {10.1093/mnras/stt064}, \href
  {https://ui.adsabs.harvard.edu/abs/2013MNRAS.430.2411M} {430, 2411}

\bibitem[\protect\citeauthoryear{{McGregor} et~al.,}{{McGregor}
  et~al.}{2003}]{mcgregor03}
{McGregor} P.~J.,  et~al., 2003, {Gemini near-infrared integral field
  spectrograph (NIFS)}.
Proceedings of the SPIE, pp 1581--1591, \mn@doi{10.1117/12.459448}

\bibitem[\protect\citeauthoryear{{Moorwood} \& {Oliva}}{{Moorwood} \&
  {Oliva}}{1990}]{moorwood90}
{Moorwood} A.~F.~M.,  {Oliva} E.,  1990, \aap, \href
  {https://ui.adsabs.harvard.edu/abs/1990A&A...239...78M} {239, 78}

\bibitem[\protect\citeauthoryear{{Mouri}}{{Mouri}}{1994}]{mouri94}
{Mouri} H.,  1994, \mn@doi [\apj] {10.1086/174184}, \href
  {https://ui.adsabs.harvard.edu/abs/1994ApJ...427..777M} {427, 777}

\bibitem[\protect\citeauthoryear{{Mouri}, {Nishida}, {Taniguchi}  \&
  {Kawara}}{{Mouri} et~al.}{1990}]{mouri90}
{Mouri} H.,  {Nishida} M.,  {Taniguchi} Y.,   {Kawara} K.,  1990, \mn@doi
  [\apj] {10.1086/169095}, \href
  {https://ui.adsabs.harvard.edu/abs/1990ApJ...360...55M} {360, 55}

\bibitem[\protect\citeauthoryear{{Mouri}, {Kawara}  \& {Taniguchi}}{{Mouri}
  et~al.}{1993}]{mouri93}
{Mouri} H.,  {Kawara} K.,   {Taniguchi} Y.,  1993, \mn@doi [\apj]
  {10.1086/172419}, \href
  {https://ui.adsabs.harvard.edu/abs/1993ApJ...406...52M} {406, 52}

\bibitem[\protect\citeauthoryear{{M{\"u}ller-S{\'a}nchez}, {Prieto}, {Hicks},
  {Vives-Arias}, {Davies}, {Malkan}, {Tacconi}  \&
  {Genzel}}{{M{\"u}ller-S{\'a}nchez} et~al.}{2011}]{ms11}
{M{\"u}ller-S{\'a}nchez} F.,  {Prieto} M.~A.,  {Hicks} E.~K.~S.,  {Vives-Arias}
  H.,  {Davies} R.~I.,  {Malkan} M.,  {Tacconi} L.~J.,   {Genzel} R.,  2011,
  \mn@doi [\apj] {10.1088/0004-637X/739/2/69}, \href
  {https://ui.adsabs.harvard.edu/abs/2011ApJ...739...69M} {739, 69}

\bibitem[\protect\citeauthoryear{{Mundell}, {Holloway}, {Pedlar}, {Meaburn},
  {Kukula}  \& {Axon}}{{Mundell} et~al.}{1995}]{mundell95}
{Mundell} C.~G.,  {Holloway} A.~J.,  {Pedlar} A.,  {Meaburn} J.,  {Kukula}
  M.~J.,   {Axon} D.~J.,  1995, \mn@doi [Monthly Notices of the Royal
  Astronomical Society] {10.1093/mnras/275.1.67}, \href
  {https://ui.adsabs.harvard.edu/abs/1995MNRAS.275...67M} {275, 67}

\bibitem[\protect\citeauthoryear{{Nagar}, {Wilson}, {Mulchaey}  \&
  {Gallimore}}{{Nagar} et~al.}{1999}]{nagar99}
{Nagar} N.~M.,  {Wilson} A.~S.,  {Mulchaey} J.~S.,   {Gallimore} J.~F.,  1999,
  \mn@doi [\apjs] {10.1086/313183}, \href
  {https://ui.adsabs.harvard.edu/abs/1999ApJS..120..209N} {120, 209}

\bibitem[\protect\citeauthoryear{{Oh} et~al.,}{{Oh} et~al.}{2018}]{BAT105}
{Oh} K.,  et~al., 2018, \mn@doi [\apjs] {10.3847/1538-4365/aaa7fd}, \href
  {https://ui.adsabs.harvard.edu/abs/2018ApJS..235....4O} {235, 4}

\bibitem[\protect\citeauthoryear{{Oliva} et~al.,}{{Oliva}
  et~al.}{2001}]{oliva01}
{Oliva} E.,  et~al., 2001, \mn@doi [\aap] {10.1051/0004-6361:20010214}, \href
  {https://ui.adsabs.harvard.edu/abs/2001A&A...369L...5O} {369, L5}

\bibitem[\protect\citeauthoryear{{Osterbrock} \& {Parker}}{{Osterbrock} \&
  {Parker}}{1965}]{osterbrock65}
{Osterbrock} D.~E.,  {Parker} R. A.~R.,  1965, \mn@doi [\apj] {10.1086/148184},
  \href {https://ui.adsabs.harvard.edu/abs/1965ApJ...141..892O} {141, 892}

\bibitem[\protect\citeauthoryear{{Prieto}, {Marco}  \& {Gallimore}}{{Prieto}
  et~al.}{2005}]{prieto05}
{Prieto} M.~A.,  {Marco} O.,   {Gallimore} J.,  2005, \mn@doi [\mnras]
  {10.1111/j.1745-3933.2005.00099.x}, \href
  {https://ui.adsabs.harvard.edu/abs/2005MNRAS.364L..28P} {364, L28}

\bibitem[\protect\citeauthoryear{{Reunanen}, {Kotilainen}  \&
  {Prieto}}{{Reunanen} et~al.}{2002}]{reunanen02}
{Reunanen} J.,  {Kotilainen} J.~K.,   {Prieto} M.~A.,  2002, \mn@doi [\mnras]
  {10.1046/j.1365-8711.2002.05181.x}, \href
  {https://ui.adsabs.harvard.edu/abs/2002MNRAS.331..154R} {331, 154}

\bibitem[\protect\citeauthoryear{{Riffel}}{{Riffel}}{2010}]{rogemar_profit}
{Riffel} R.~A.,  2010, \mn@doi [\apss] {10.1007/s10509-010-0317-y}, \href
  {https://ui.adsabs.harvard.edu/abs/2010Ap&SS.327..239R} {327, 239}

\bibitem[\protect\citeauthoryear{{Riffel} \& {Storchi-Bergmann}}{{Riffel} \&
  {Storchi-Bergmann}}{2011a}]{rogemar_mrk1066_kin}
{Riffel} R.~A.,  {Storchi-Bergmann} T.,  2011a, \mn@doi [\mnras]
  {10.1111/j.1365-2966.2010.17721.x}, \href
  {https://ui.adsabs.harvard.edu/abs/2011MNRAS.411..469R} {411, 469}

\bibitem[\protect\citeauthoryear{{Riffel} \& {Storchi-Bergmann}}{{Riffel} \&
  {Storchi-Bergmann}}{2011b}]{rogemar_mrk1157}
{Riffel} R.~A.,  {Storchi-Bergmann} T.,  2011b, \mn@doi [\mnras]
  {10.1111/j.1365-2966.2011.19441.x}, \href
  {https://ui.adsabs.harvard.edu/abs/2011MNRAS.417.2752R} {417, 2752}

\bibitem[\protect\citeauthoryear{{Riffel}, {Storchi-Bergmann}, {Winge}  \&
  {Barbosa}}{{Riffel} et~al.}{2006a}]{rogemar_eso428}
{Riffel} R.~A.,  {Storchi-Bergmann} T.,  {Winge} C.,   {Barbosa} F. K.~B.,
  2006a, \mn@doi [\mnras] {10.1111/j.1365-2966.2006.11050.x}, \href
  {https://ui.adsabs.harvard.edu/abs/2006MNRAS.373....2R} {373, 2}

\bibitem[\protect\citeauthoryear{{Riffel}, {Rodr{\'\i}guez-Ardila}  \&
  {Pastoriza}}{{Riffel} et~al.}{2006b}]{rogerio06}
{Riffel} R.,  {Rodr{\'\i}guez-Ardila} A.,   {Pastoriza} M.~G.,  2006b, \mn@doi
  [\aap] {10.1051/0004-6361:20065291}, \href
  {https://ui.adsabs.harvard.edu/abs/2006A&A...457...61R} {457, 61}

\bibitem[\protect\citeauthoryear{{Riffel}, {Storchi-Bergmann}, {Winge},
  {McGregor}, {Beck}  \& {Schmitt}}{{Riffel} et~al.}{2008}]{rogemarN4051}
{Riffel} R.~A.,  {Storchi-Bergmann} T.,  {Winge} C.,  {McGregor} P.~J.,  {Beck}
  T.,   {Schmitt} H.,  2008, \mn@doi [\mnras]
  {10.1111/j.1365-2966.2008.12936.x}, \href
  {https://ui.adsabs.harvard.edu/abs/2008MNRAS.385.1129R} {385, 1129}

\bibitem[\protect\citeauthoryear{{Riffel}, {Storchi-Bergmann}  \&
  {Nagar}}{{Riffel} et~al.}{2010}]{rogemarMrk1066exc}
{Riffel} R.~A.,  {Storchi-Bergmann} T.,   {Nagar} N.~M.,  2010, \mn@doi
  [\mnras] {10.1111/j.1365-2966.2010.16308.x}, \href
  {https://ui.adsabs.harvard.edu/abs/2010MNRAS.404..166R} {404, 166}

\bibitem[\protect\citeauthoryear{{Riffel}, {Rodr{\'\i}guez-Ardila}, {Aleman},
  {Brotherton}, {Pastoriza}, {Bonatto}  \& {Dors}}{{Riffel}
  et~al.}{2013a}]{rogerio13}
{Riffel} R.,  {Rodr{\'\i}guez-Ardila} A.,  {Aleman} I.,  {Brotherton} M.~S.,
  {Pastoriza} M.~G.,  {Bonatto} C.,   {Dors} O.~L.,  2013a, \mn@doi [\mnras]
  {10.1093/mnras/stt026}, \href
  {https://ui.adsabs.harvard.edu/abs/2013MNRAS.430.2002R} {430, 2002}

\bibitem[\protect\citeauthoryear{{Riffel}, {Storchi-Bergmann}  \&
  {Winge}}{{Riffel} et~al.}{2013b}]{rogemar_mrk79}
{Riffel} R.~A.,  {Storchi-Bergmann} T.,   {Winge} C.,  2013b, \mn@doi [\mnras]
  {10.1093/mnras/stt045}, \href
  {https://ui.adsabs.harvard.edu/abs/2013MNRAS.430.2249R} {430, 2249}

\bibitem[\protect\citeauthoryear{{Riffel}, {Vale}, {Storchi-Bergmann}  \&
  {McGregor}}{{Riffel} et~al.}{2014a}]{rogemar_n1068}
{Riffel} R.~A.,  {Vale} T.~B.,  {Storchi-Bergmann} T.,   {McGregor} P.~J.,
  2014a, \mn@doi [\mnras] {10.1093/mnras/stu843}, \href
  {https://ui.adsabs.harvard.edu/abs/2014MNRAS.442..656R} {442, 656}

\bibitem[\protect\citeauthoryear{{Riffel}, {Storchi-Bergmann}  \&
  {Riffel}}{{Riffel} et~al.}{2014b}]{rogemar_n5929_let}
{Riffel} R.~A.,  {Storchi-Bergmann} T.,   {Riffel} R.,  2014b, \mn@doi [\apjl]
  {10.1088/2041-8205/780/2/L24}, \href
  {https://ui.adsabs.harvard.edu/abs/2014ApJ...780L..24R} {780, L24}

\bibitem[\protect\citeauthoryear{{Riffel}, {Storchi-Bergmann}  \&
  {Riffel}}{{Riffel} et~al.}{2015}]{rogemar_n5929}
{Riffel} R.~A.,  {Storchi-Bergmann} T.,   {Riffel} R.,  2015, \mn@doi [\mnras]
  {10.1093/mnras/stv1129}, \href
  {https://ui.adsabs.harvard.edu/abs/2015MNRAS.451.3587R} {451, 3587}

\bibitem[\protect\citeauthoryear{{Riffel}, {Storchi-Bergmann}, {Riffel},
  {Dahmer-Hahn}, {Diniz}, {Sch{\"o}nell}  \& {Dametto}}{{Riffel}
  et~al.}{2017}]{rogemar_stellar}
{Riffel} R.~A.,  {Storchi-Bergmann} T.,  {Riffel} R.,  {Dahmer-Hahn} L.~G.,
  {Diniz} M.~R.,  {Sch{\"o}nell} A.~J.,   {Dametto} N.~Z.,  2017, \mn@doi
  [\mnras] {10.1093/mnras/stx1308}, \href
  {https://ui.adsabs.harvard.edu/abs/2017MNRAS.470..992R} {470, 992}

\bibitem[\protect\citeauthoryear{{Riffel} et~al.,}{{Riffel}
  et~al.}{2018}]{rogemar_sample}
{Riffel} R.~A.,  et~al., 2018, \mn@doi [\mnras] {10.1093/mnras/stx2857}, \href
  {https://ui.adsabs.harvard.edu/abs/2018MNRAS.474.1373R} {474, 1373}

\bibitem[\protect\citeauthoryear{{Riffel} et~al.,}{{Riffel}
  et~al.}{2019}]{rogerio19}
{Riffel} R.,  et~al., 2019, \mn@doi [\mnras] {10.1093/mnras/stz1077}, \href
  {https://ui.adsabs.harvard.edu/abs/2019MNRAS.486.3228R} {486, 3228}

\bibitem[\protect\citeauthoryear{{Riffel} et~al.,}{{Riffel}
  et~al.}{2020a}]{rogemar20_letter}
{Riffel} R.~A.,  et~al., 2020a, \mn@doi [\mnras] {10.1093/mnrasl/slaa194},
  \href {https://ui.adsabs.harvard.edu/abs/2020MNRAS.tmpL.235R} {}

\bibitem[\protect\citeauthoryear{{Riffel}, {Storchi-Bergmann}, {Zakamska}  \&
  {Riffel}}{{Riffel} et~al.}{2020b}]{rogemar_N1275}
{Riffel} R.~A.,  {Storchi-Bergmann} T.,  {Zakamska} N.~L.,   {Riffel} R.,
  2020b, \mn@doi [\mnras] {10.1093/mnras/staa1922}, \href
  {https://ui.adsabs.harvard.edu/abs/2020MNRAS.tmp.2053R} {496, 4857}

\bibitem[\protect\citeauthoryear{{Rodr{\'\i}guez-Ardila} \&
  {Fonseca-Faria}}{{Rodr{\'\i}guez-Ardila} \& {Fonseca-Faria}}{2020}]{ardila20}
{Rodr{\'\i}guez-Ardila} A.,  {Fonseca-Faria} M.~A.,  2020, \mn@doi [\apjl]
  {10.3847/2041-8213/ab901b}, \href
  {https://ui.adsabs.harvard.edu/abs/2020ApJ...895L...9R} {895, L9}

\bibitem[\protect\citeauthoryear{{Rodr{\'\i}guez-Ardila}, {Viegas}, {Pastoriza}
   \& {Prato}}{{Rodr{\'\i}guez-Ardila} et~al.}{2002}]{ardila02}
{Rodr{\'\i}guez-Ardila} A.,  {Viegas} S.~M.,  {Pastoriza} M.~G.,   {Prato} L.,
  2002, \mn@doi [\apj] {10.1086/342840}, \href
  {https://ui.adsabs.harvard.edu/abs/2002ApJ...579..214R} {579, 214}

\bibitem[\protect\citeauthoryear{{Rodr{\'\i}guez-Ardila}, {Pastoriza},
  {Viegas}, {Sigut}  \& {Pradhan}}{{Rodr{\'\i}guez-Ardila}
  et~al.}{2004}]{ardila04}
{Rodr{\'\i}guez-Ardila} A.,  {Pastoriza} M.~G.,  {Viegas} S.,  {Sigut}
  T.~A.~A.,   {Pradhan} A.~K.,  2004, \mn@doi [\aap]
  {10.1051/0004-6361:20034285}, \href
  {https://ui.adsabs.harvard.edu/abs/2004A&A...425..457R} {425, 457}

\bibitem[\protect\citeauthoryear{{Rodr{\'\i}guez-Ardila}, {Riffel}  \&
  {Pastoriza}}{{Rodr{\'\i}guez-Ardila} et~al.}{2005}]{ardila05}
{Rodr{\'\i}guez-Ardila} A.,  {Riffel} R.,   {Pastoriza} M.~G.,  2005, \mn@doi
  [\mnras] {10.1111/j.1365-2966.2005.09638.x}, \href
  {https://ui.adsabs.harvard.edu/abs/2005MNRAS.364.1041R} {364, 1041}

\bibitem[\protect\citeauthoryear{{Rodr{\'\i}guez-Ardila}, {Prieto}, {Viegas}
  \& {Gruenwald}}{{Rodr{\'\i}guez-Ardila} et~al.}{2006}]{ardila06}
{Rodr{\'\i}guez-Ardila} A.,  {Prieto} M.~A.,  {Viegas} S.,   {Gruenwald} R.,
  2006, \mn@doi [\apj] {10.1086/508864}, \href
  {https://ui.adsabs.harvard.edu/abs/2006ApJ...653.1098R} {653, 1098}

\bibitem[\protect\citeauthoryear{{Rodr{\'\i}guez-Ardila}, {Prieto}, {Portilla}
  \& {Tejeiro}}{{Rodr{\'\i}guez-Ardila} et~al.}{2011}]{ardila11}
{Rodr{\'\i}guez-Ardila} A.,  {Prieto} M.~A.,  {Portilla} J.~G.,   {Tejeiro}
  J.~M.,  2011, \mn@doi [\apj] {10.1088/0004-637X/743/2/100}, \href
  {https://ui.adsabs.harvard.edu/abs/2011ApJ...743..100R} {743, 100}

\bibitem[\protect\citeauthoryear{{Rodr{\'\i}guez-Ardila}, {Prieto}, {Mazzalay},
  {Fern{\'a}ndez-Ontiveros}, {Luque}  \&
  {M{\"u}ller-S{\'a}nchez}}{{Rodr{\'\i}guez-Ardila} et~al.}{2017}]{ardila17}
{Rodr{\'\i}guez-Ardila} A.,  {Prieto} M.~A.,  {Mazzalay} X.,
  {Fern{\'a}ndez-Ontiveros} J.~A.,  {Luque} R.,   {M{\"u}ller-S{\'a}nchez} F.,
  2017, \mn@doi [\mnras] {10.1093/mnras/stx1401}, \href
  {https://ui.adsabs.harvard.edu/abs/2017MNRAS.470.2845R} {470, 2845}

\bibitem[\protect\citeauthoryear{Ruschel-Dutra}{Ruschel-Dutra}{2020}]{ifscube}
Ruschel-Dutra D.,  2020, danielrd6/ifscube v1.0,
  \mn@doi{10.5281/zenodo.3945237}, \url
  {https://doi.org/10.5281/zenodo.3945237}

\bibitem[\protect\citeauthoryear{{Schinnerer}, {Eckart}  \&
  {Tacconi}}{{Schinnerer} et~al.}{2000}]{schinnerer00}
{Schinnerer} E.,  {Eckart} A.,   {Tacconi} L.~J.,  2000, \mn@doi [The
  Astrophysical Journal] {10.1086/308703}, \href
  {https://ui.adsabs.harvard.edu/abs/2000ApJ...533..826S} {533, 826}

\bibitem[\protect\citeauthoryear{{Schmitt}, {Ulvestad}, {Antonucci}  \&
  {Kinney}}{{Schmitt} et~al.}{2001}]{schmitt01}
{Schmitt} H.~R.,  {Ulvestad} J.~S.,  {Antonucci} R.~R.~J.,   {Kinney} A.~L.,
  2001, \mn@doi [\apjs] {10.1086/318957}, \href
  {https://ui.adsabs.harvard.edu/abs/2001ApJS..132..199S} {132, 199}

\bibitem[\protect\citeauthoryear{{Schmitt}, {Donley}, {Antonucci}, {Hutchings}
  \& {Kinney}}{{Schmitt} et~al.}{2003}]{schmitt03}
{Schmitt} H.~R.,  {Donley} J.~L.,  {Antonucci} R.~R.~J.,  {Hutchings} J.~B.,
  {Kinney} A.~L.,  2003, \mn@doi [\apjs] {10.1086/377440}, \href
  {https://ui.adsabs.harvard.edu/abs/2003ApJS..148..327S} {148, 327}

\bibitem[\protect\citeauthoryear{{Sch{\"o}nell}, {Storchi-Bergmann}, {Riffel},
  {Riffel}, {Bianchin}, {Dahmer-Hahn}, {Diniz}  \& {Dametto}}{{Sch{\"o}nell}
  et~al.}{2019}]{astor19}
{Sch{\"o}nell} A.~J.,  {Storchi-Bergmann} T.,  {Riffel} R.~A.,  {Riffel} R.,
  {Bianchin} M.,  {Dahmer-Hahn} L.~G.,  {Diniz} M.~R.,   {Dametto} N.~Z.,
  2019, \mn@doi [\mnras] {10.1093/mnras/stz523}, \href
  {https://ui.adsabs.harvard.edu/abs/2019MNRAS.485.2054S} {485, 2054}

\bibitem[\protect\citeauthoryear{{Shields} \& {Oke}}{{Shields} \&
  {Oke}}{1975}]{shields75}
{Shields} G.~A.,  {Oke} J.~B.,  1975, \mn@doi [\apj] {10.1086/153482}, \href
  {https://ui.adsabs.harvard.edu/abs/1975ApJ...197....5S} {197, 5}

\bibitem[\protect\citeauthoryear{{Simpson}, {Forbes}, {Baker}  \&
  {Ward}}{{Simpson} et~al.}{1996}]{simpson96}
{Simpson} C.,  {Forbes} D.~A.,  {Baker} A.~C.,   {Ward} M.~J.,  1996, \mn@doi
  [\mnras] {10.1093/mnras/283.3.777}, \href
  {https://ui.adsabs.harvard.edu/abs/1996MNRAS.283..777S} {283, 777}

\bibitem[\protect\citeauthoryear{{Smith}}{{Smith}}{1995}]{Smith95}
{Smith} M.~D.,  1995, \aap, \href
  {https://ui.adsabs.harvard.edu/abs/1995A&A...296..789S} {296, 789}

\bibitem[\protect\citeauthoryear{{Sternberg} \& {Dalgarno}}{{Sternberg} \&
  {Dalgarno}}{1989}]{Sternberg89}
{Sternberg} A.,  {Dalgarno} A.,  1989, \mn@doi [\apj] {10.1086/167193}, \href
  {https://ui.adsabs.harvard.edu/abs/1989ApJ...338..197S} {338, 197}

\bibitem[\protect\citeauthoryear{{Storchi-Bergmann}, {McGregor}, {Riffel},
  {Sim{\~o}es Lopes}, {Beck}  \& {Dopita}}{{Storchi-Bergmann}
  et~al.}{2009}]{sbN4151Exc}
{Storchi-Bergmann} T.,  {McGregor} P.~J.,  {Riffel} R.~A.,  {Sim{\~o}es Lopes}
  R.,  {Beck} T.,   {Dopita} M.,  2009, \mn@doi [\mnras]
  {10.1111/j.1365-2966.2009.14388.x}, \href
  {https://ui.adsabs.harvard.edu/abs/2009MNRAS.394.1148S} {394, 1148}

\bibitem[\protect\citeauthoryear{{Storchi-Bergmann}, {Lopes}, {McGregor},
  {Riffel}, {Beck}  \& {Martini}}{{Storchi-Bergmann}
  et~al.}{2010}]{sb_N4151_kin}
{Storchi-Bergmann} T.,  {Lopes} R.~D.~S.,  {McGregor} P.~J.,  {Riffel} R.~A.,
  {Beck} T.,   {Martini} P.,  2010, \mn@doi [\mnras]
  {10.1111/j.1365-2966.2009.15962.x}, \href
  {https://ui.adsabs.harvard.edu/abs/2010MNRAS.402..819S} {402, 819}

\bibitem[\protect\citeauthoryear{{Turner}, {Kirby-Docken}  \&
  {Dalgarno}}{{Turner} et~al.}{1977}]{turner77}
{Turner} J.,  {Kirby-Docken} K.,   {Dalgarno} A.,  1977, \mn@doi [\apjs]
  {10.1086/190481}, \href
  {https://ui.adsabs.harvard.edu/abs/1977ApJS...35..281T} {35, 281}

\bibitem[\protect\citeauthoryear{{Veilleux}, {Tully}  \&
  {Bland-Hawthorn}}{{Veilleux} et~al.}{1993}]{veilleux93}
{Veilleux} S.,  {Tully} R.~B.,   {Bland-Hawthorn} J.,  1993, \mn@doi
  [Astronomical Journal] {10.1086/116512}, \href
  {https://ui.adsabs.harvard.edu/abs/1993AJ....105.1318V} {105, 1318}

\bibitem[\protect\citeauthoryear{{Veilleux}, {Goodrich}  \& {Hill}}{{Veilleux}
  et~al.}{1997}]{veilleux97}
{Veilleux} S.,  {Goodrich} R.~W.,   {Hill} G.~J.,  1997, \mn@doi [\apj]
  {10.1086/303735}, \href
  {https://ui.adsabs.harvard.edu/abs/1997ApJ...477..631V} {477, 631}

\bibitem[\protect\citeauthoryear{{V{\'e}ron-Cetty} \&
  {V{\'e}ron}}{{V{\'e}ron-Cetty} \& {V{\'e}ron}}{2006}]{veron-cetty06}
{V{\'e}ron-Cetty} M.~P.,  {V{\'e}ron} P.,  2006, \mn@doi [Astronomy and
  Astrophysics] {10.1051/0004-6361:20065177}, \href
  {https://ui.adsabs.harvard.edu/abs/2006A&A...455..773V} {455, 773}

\bibitem[\protect\citeauthoryear{{Viegas-Aldrovandi} \&
  {Contini}}{{Viegas-Aldrovandi} \& {Contini}}{1989}]{viegas89}
{Viegas-Aldrovandi} S.~M.,  {Contini} M.,  1989, \aap, \href
  {https://ui.adsabs.harvard.edu/abs/1989A&A...215..253V} {215, 253}

\bibitem[\protect\citeauthoryear{{Wylezalek} et~al.,}{{Wylezalek}
  et~al.}{2017}]{wylezalek17}
{Wylezalek} D.,  et~al., 2017, \mn@doi [\mnras] {10.1093/mnras/stx246}, \href
  {https://ui.adsabs.harvard.edu/abs/2017MNRAS.467.2612W} {467, 2612}

\bibitem[\protect\citeauthoryear{{Wylezalek}, {Flores}, {Zakamska}, {Greene}
  \& {Riffel}}{{Wylezalek} et~al.}{2020}]{wylezalek20}
{Wylezalek} D.,  {Flores} A.~M.,  {Zakamska} N.~L.,  {Greene} J.~E.,   {Riffel}
  R.~A.,  2020, \mn@doi [\mnras] {10.1093/mnras/staa062}, \href
  {https://ui.adsabs.harvard.edu/abs/2020MNRAS.492.4680W} {492, 4680}

\bibitem[\protect\citeauthoryear{{Zakamska} \& {Greene}}{{Zakamska} \&
  {Greene}}{2014}]{zakamska14}
{Zakamska} N.~L.,  {Greene} J.~E.,  2014, \mn@doi [\mnras]
  {10.1093/mnras/stu842}, \href
  {https://ui.adsabs.harvard.edu/abs/2014MNRAS.442..784Z} {442, 784}

\bibitem[\protect\citeauthoryear{{de Vaucouleurs}, {de Vaucouleurs}, {Corwin},
  {Buta}, {Paturel}  \& {Fouque}}{{de Vaucouleurs}
  et~al.}{1991}]{devaucouleurs91}
{de Vaucouleurs} G.,  {de Vaucouleurs} A.,  {Corwin} Herold~G. J.,  {Buta}
  R.~J.,  {Paturel} G.,   {Fouque} P.,  1991, {Third Reference Catalogue of
  Bright Galaxies}

\makeatother
\end{thebibliography}

% Don't change these lines
\bsp	% typesetting comment
\label{lastpage}
\end{document}